\documentclass{rspublic}
 
\usepackage{epsf}
\usepackage{longtable}
\usepackage{natbib}
\usepackage{epstopdf}
\usepackage{graphicx}

\def\aap{Astron. Astrophys.}
\def\apj{Astrophys. J.}
\def\apjl{Astrophys. J. Lett.}
\def\solphys{Solar Phys.}
\def\ssr{Space~Sci.~Rev.}
\def\pasj{PASJ}
\def\prl{Physical Review Letters}

\def\mnras{Montly Notices of the Royal Astronomical Society}
\def\nat{Nature}
\def\araa{Annual Review of Astronomy and Astrophysics}
\def\zap{Zeitschrift f\"{u}r Astrophysik}

\begin{document} 
 
\title[Wave Heating of the Solar Atmosphere]{Wave Heating of the Solar Atmosphere} 
 
\author[I\~nigo Arregui]{I\~nigo Arregui} 
 
\affiliation{Instituto de Astrof\'{\i}sica de Canarias, V\'{\i}a L\'actea s/n, E-38205 La Laguna, Tenerife, Spain\\
Departamento de Astrof\'{\i}sica, Universidad de La Laguna, E-38206 La Laguna, Tenerife, Spain} 
 
\maketitle 
 
\begin{abstract}{Sun, Magnetic Fields, Magnetohydrodynamics (MHD), Waves, Coronal Heating} 
Magnetic waves are a relevant component in the dynamics of the solar atmosphere. Their significance has increased because of their potential as a remote diagnostic tool and their presumed contribution to plasma heating processes. We discuss our current understanding on coronal heating by 
magnetic waves, based on recent observational evidence and theoretical advances. The discussion starts with a selection of observational discoveries that have brought magnetic waves to the forefront of the coronal heating discussion. Then, our theoretical understanding on the nature and properties of the observed waves and the physical processes that have been proposed to explain observations are described. Particular attention is given to the sequence of processes that link observed wave characteristics with concealed energy transport, dissipation, and heat conversion. We conclude with a commentary on how the combination of theory and observations should help us understanding and quantifying magnetic wave heating of the solar atmosphere. 
\end{abstract} 

%%%%%%%%%%%%%%%%%%%%%%%%%%%%%%%%%%%%%%%%%%%%%%%%%%%%%%%%%%%%%%%%%% 
\section{Introduction}
%%%%%%%%%%%%%%%%%%%%%%%%%%%%%%%%%%%%%%%%%%%%%%%%%%%%%%%%%%%%%%%%%% 
In spite of decades of advances in observations and theory, the solar coronal heating problem remains unsolved. 
A full understanding about the magnetic and plasma structuring of the corona, the nature of the plasma dynamics, and the physical processes that create and maintain a hot atmosphere remain elusive \citep[see][for reviews]{kuperus69,withbroe77,kuperus81,Narain96,aschwanden05b,klimchuk06,hood10,parnell12}.

The coronal heating problem originated more than seventy years ago, when \cite{grotrian39} and \cite{edlen43} identified the presence of Fe IX and Ca XIV spectral lines in the light emitted by the corona,
indicating the presence of fully ionised plasma at multi-million degrees. In a recent historical note, \cite{peter14}  assign the merit to \cite{alfven41}, who summarised six arguments to support that the corona is heated to such high temperatures. Since then, solar physics research has sought to identify the physical mechanism(s) that might balance the thermal conduction, radiation, and solar wind losses, first quantified by \cite{withbroe77}. Researchers in the field agree upon the fact that the origin for such temperatures is of magnetic nature and that the energy source lies in the solar surface plasma motions. This available energy is then transported to the upper layers of the solar atmosphere where it gets dissipated. The exact physical processes and their contribution to the heating of the plasma remain largely unknown or unquantified \citep[][]{parnell12}. 

A number of mechanisms are believed to contribute to the heating: the direct dissipation of magnetic energy by processes such as magnetic reconnection \citep{sturrock81}, current cascades \citep{parker63}, viscous turbulence \citep{vanballegooijen86}, or magnetic field braiding \citep{peter04}; the dissipation of magnetic wave energy stored, transported, and transferred to small scales by waves \citep{alfven47,ionson78,heyvaerts83,goossens91,hood97}; or the mass flow cycle between chromosphere and corona \citep{mcintosh12} are some suggested mechanisms. There is ample evidence for the occurrence of magnetic reconnection processes \citep{zweibel09}; magnetic wave dynamics \citep{nakariakov05}; and mass flows connecting chromospheric and coronal regions \citep{peter99}. All these processes are under theoretical and observational study and advances on their description have been made so a discussion in terms of mutually exclusive mechanisms is beside the point. It is important however to assess the importance of each mechanism. 

Wave energy transport and dissipation were regarded as relevant soon after the coronal heating problem arose. The first proposed theories involved wave-based mechanisms.  \cite{biermann46} and \cite{schwarzschild48} suggested that acoustic waves, generated in the convection zone, supplied the non-radiative energy needed to heat the chromosphere and corona in the form of mechanical energy transported outwards \citep[see review by][]{stein74}. Magnetic wave heating  was proposed by \cite{alfven47} soon after the existence of magnetohydrodynamic (MHD) waves was postulated by himself \citep{alfven42}.  It was soon realised that the dissipation of magnetic waves could have an important role in coronal heating, because surface Alfv\'en waves can be transmitted upwards in the atmosphere \citep{hollweg78} and heat the corona \citep{wentzel74}. These waves are indeed subject to mechanisms, such as resonant absorption \citep{ionson78} and phase-mixing \citep{heyvaerts83}, that enhance the dissipation of energy by resistive or viscous processes, which would otherwise take too long in the high Reynolds number corona. Theoretical and numerical studies on the dissipation of hydromagnetic waves \citep{wentzel78,wentzel79}  and the resonance absorption of Alfv\'en waves \citep{ionson78,rae81,poedts89,poedts90} were carried out in open and closed magnetic structures. Numerical experiments included the sideways or footpoint driving of the structures, trying to mimic the buffeting of the solar surface \citep{goedbloed94,ofman95}. The first relations to loop observations were made by \cite{hollweg84}, to explain data by \cite{golub80}. However, proper instrumentation able to detect waves and reliably measure their properties was not available.

This changed near the end of the last century, when routine observations of the solar atmosphere made with instruments onboard the Solar and Heliospheric Observatory (SoHO) and the Transition Region and Coronal Explorer (TRACE)  led to  a number of discoveries. Remarkable examples are the transverse oscillations in coronal loops \citep{aschwanden99,nakariakov99}  and their rapid time damping; the compressive waves in short and long coronal loops \citep{berghmans99, nightingale99,demoortel00}; the quasi-periodic oscillations of the Doppler shift and intensity in emission
lines in coronal loops \citep{kliem02,wang02a} and their damping;  or the quasi-periodic disturbances in open magnetic regions of the corona \citep{ofman97,deforest98}. Extended reviews on these observations can be found in \cite{nakariakov05}, \cite{demoortel05}, \cite{aschwanden06}, and \cite{demoortel12}.

The observed wave activity is interpreted in terms of standing and propagating MHD waves. Nowadays, the existence of theoretically predicted MHD wave types in the solar atmosphere is well established \citep{roberts00, aschwanden03a,nakariakov05,ballester06}. This has lead to the development of solar atmospheric seismology \citep{uchida70,roberts84}, which enables the determination of physical parameters in the solar atmosphere from the comparison of observed and theoretical wave properties. The application of this method to the solar atmosphere has been successful, but the wave heating perspective was discontinued for more than a decade. 

A renewed interest has now grown on the wave heating question. This was motivated by the evidence for ubiquitous wave activity at different layers of the solar atmosphere, gathered by observations with instruments onboard e.g., the Hinode and Solar Dynamics Observatory (SDO) spacecraft or the Hi-C rocket experiment. These high resolution observations have demonstrated the presence of oscillations of magnetic and plasma structures in solar chromospheric spicules \citep{depontieu07}; coronal magnetic loops \citep{mcintosh11} prominence plasmas \citep{okamoto07,arregui12a}; X-ray jets \citep{cirtain07}; and extended regions of the corona \citep{tomczyk07,tomczyk09}. The observed disturbances display time and spatial damping possibly indicating in situ energy dissipation. In some cases, their energy content seems to be large enough to compensate estimated energy losses. These observations have replaced coronal wave heating at the forefront of the discussion, leading to the need to reassess the significance of wave energy transport and dissipation to the heating problem \citep{cargill11,parnell12}. 

We discuss the role of magnetic wave energy transport and dissipation processes in the heating of the solar corona, based on our current theoretical understanding and observational evidence. We focus on the following aspects: in Section \ref{observations}, we present a selection of wave activity observations by pointing out their relevance to the heating  problem; Section \ref{theory} offers a description of theoretical models and physical processes that have been suggested to explain observations; particular emphasis is made in Section \ref{processes} on the series of physical processes that link wave energy content, propagation, damping and dissipation. Section \ref{datatheory} contains a commentary on how the combination of data and theory, by means of forward modelling of observational signatures and seismic inversion of observed wave properties, should help advance the field in the future. Our conclusions are presented in Section \ref{conclusions}.  

%%%%%%%%%%%%%%%%%%%%%%%%%%%%%%%%%%%%%%%%%%%%%%%%%%%%%%%%%%%%%%%%%% 
\section{Observed Wave Activity}\label{observations}
%%%%%%%%%%%%%%%%%%%%%%%%%%%%%%%%%%%%%%%%%%%%%%%%%%%%%%%%%%%%%%%%%% 

Early observations already pointed to the existence of quasi-periodic perturbations in solar coronal structures  \citep[see e.g.,][]{billings59,rosenberg70,vernazza75,tsubaki77,trottet79, antonucci84,aschwanden87,deubner89}. The detection of these oscillations was based on the measurement of the temporal and spatial variation of spectroscopic properties (intensity, width, and Doppler velocity) of coronal emission lines. Some of these observations took advantage of the instrumentation onboard Skylab or Yohkoh spacecraft, but were mostly restricted to time series analyses with little spatial information \citep[see][for a summary of early observations]{tsubaki88a}. 

This situation changed with the advent of the imaging and spectroscopic instruments onboard SoHO and TRACE spacecraft. Due to the temperature discrimination and spatial resolution capabilities of their EUV and soft X-ray telescopes, observations demonstrated the existence of wave-like dynamics in the form of e.g., density fluctuations in plumes and coronal holes \citep{ofman97,deforest98,ofman00a}, propagating compressional disturbances in coronal loops \citep{nightingale99,berghmans99,demoortel00,robbrecht01,demoortel02c,demoortel02d}, or transverse coronal loop oscillations \citep{aschwanden99,nakariakov99}.

The existence of waves and oscillations in magnetic and plasma structures of the solar atmosphere is now beyond question. The spatial, temporal, and spectral resolution of imaging and spectroscopic instruments in current ground- and space-based observatories (SST, DST, SoHO, TRACE CoMP, Hinode, STEREO, SDO, HI-C, IRIS) have enabled us to directly image and measure motions associated to wave dynamics with increasing precision. Data show evidence for waves in increasingly finer plasma structures belonging to regions of the solar atmosphere with different physical properties, such as coronal loops; prominence plasmas; chromospheric spicules and mottles; coronal holes, plumes, etc. Let us focus on some particular examples and draw attention to the aspects that are relevant to wave heating.

Transverse loop oscillations offered the first imaging evidence for standing waves in the corona associated with the periodic displacement of these structures \citep{aschwanden99,nakariakov99,aschwanden02,schrijver02}. In these events wave damping can be directly observed and measured.  This constitutes important information for seismology diagnostics \citep{arregui07a,pascoe13}  and the reason for considering damping mechanisms such as resonant absorption \citep{goossens06}. However, TRACE data showed that these were occasional events and loops are hot irrespective of the presence or absence of lateral displacements. On the other hand, the Coronal Multi-channel Polarimeter (CoMP) observations indicate that coronal disturbances are present in extended regions of the corona (1.05--1.35 R$_{\rm Sun}$). The measured Doppler velocity fluctuations, of the order of 0.3 km s$^{-1}$,  do not produce significant intensity variations and are interpreted as Alfv\'en waves propagating along the coronal magnetic field, with  measured phase speeds of the order of 1 to 4 Mm s$^{-1}$ \citep{tomczyk07,tomczyk09}.  Importantly, CoMP data show signatures of in situ wave damping in the form of a discrepancy in the outward to inward wave power, but energy estimates ($\sim$ 0.01 W m$^{-1}$) seem to fall below the amount required to heat the ambient plasma. The Solar Dynamics Observatory (SDO) has shown us that Alfv\'en waves are common in the transition regions and corona \citep{mcintosh11} and that motions visible in both regions share a common origin \citep{depontieu11}. Energy estimates vary from one region to another. The disturbances seem to be energetic enough to power the quiet Sun corona and coronal hole regions, with an estimated wave power of $\sim$ 100 -- 200 W m$^{-2}$, but not the active region corona ($\sim$100 W m$^{-2}$ in front of the required 2,000 W m$^{-2}$). Observations of transverse waves in the active region corona by \cite{morton13a}, obtained by combining the highest available resolution instruments onboard Hi-C and SDO/AIA, confirm that wave activity is all-pervasive in the EUV corona. However, this activity is concluded to be of low energy because of the relatively low displacement and velocity amplitudes ($<$ 50 km and $<$ 3 km s$^{-1}$, respectively) in the measured waves. 

Chromospheric spicules are excellent candidates for mass and wave energy  transport to the corona, since they provide a physical connection between both media. They are present everywhere and at all times and show a complex dynamics with a combination of lateral swaying, upward mass flows, and torsional motions \citep{okamoto11,depontieu12}. Recent observations by \cite{morton14} suggest that wave damping is  present in these structures. The recent analyses by the IRIS team will clarify many aspects about their field and plasma structuring and dynamics \citep{pereira14,skogsrud14}. Propagating and standing transverse oscillations are also present in chromospheric mottles belonging to strong photospheric field concentrations. By measuring the amplitude and phase speed variation along these structures, \cite{kuridze13} suggest that their sudden change at a given height is an indication of the presence of non-linear wave processes. Photospheric field concentrations, such as bright points, were proposed as the source of torsional Alfv\'en waves by \cite{jess09}. The increasing evidence for surface vortex motions \citep{bonet08,wedemeyer09} strongly suggests this type of dynamics can be important in the generation and wave energy propagation to  upper layers \citep{mathioudakis13}.  The chromosphere seems to be a vast reservoir of MHD wave energy with observations showing a complex combination of intensity and width variations of chromospheric structures together with their transverse displacement \citep{morton12}. The evidence points to the photospheric excitation of the dynamics  observed at the upper layers \citep{morton13b}

Waves in polar coronal hole regions are gaining ample attention. The measurement of wave characteristics, such as the amplitude, periodicity, and phase speed provide plasma and field diagnostics and constraints on theoretical models of coronal heating and solar wind acceleration \citep{banerjee11}. Recent analyses have enabled the measurement of the spatial variation of the non-thermal velocity for Alfv\'en waves and the electron density along the propagation direction. \cite{banerjee09} find that the non-thermal velocity is inversely proportional to the quadratic root of the electron density, in agreement with what is predicted for undamped radially propagating linear Alfv\'en waves. \cite{hahn12}  report that the widths of spectral lines, which are assumed to be proportional to the wave amplitude, decrease at relatively low heights. This  is interpreted as an indication of wave damping. The energy dissipated in between 1 and 1.3 solar radii is estimated to account for up to 70\% of the amount required to heat the polar coronal hole and accelerate the solar wind.  Similar analyses have quantified the energy carried out and dissipated by Alfv\'en waves in these structures \citep{bemporad12,hahn13}.  By combining spectroscopic measurements with a magnetic field model, \cite{hahn14} were able to trace the variation of the non thermal velocity along the magnetic field above the limb of a quiet sun region. They find that the waves are dissipated over a region centred on the top of the loops. The position along the loop where the damping begins  is strongly correlated with the length of the loop, implying that the damping mechanism depends on the global loop properties rather than on local collisional dissipation. Transverse waves seem to be present also in solar plumes and  \cite{thurgood14} have measured the occurrence rate of wave properties, such as periods and displacement and velocity amplitudes in great detail. 

The relationship between wave signatures at different heights of the solar atmosphere is reported in many studies \citep[see e.g., ][]{oshea02,brynildsen03,marsh06}. They provide information about the phase speed of propagating waves from the time delay of signatures detected at different heights in the solar atmosphere \citep{abramov11}.  The high spatial and temporal resolution of SDO/AIA is enabling additional analyses of the spatial distribution of the frequency, time series analyses, and correlations between the signals at different atmospheric heights \citep{reznikova12,freij14}. However, obtaining a causal connection between wave dynamics at different regions of the atmosphere is still problematic. Understanding how the waves are generated and behave as a function of the line formation temperature and the magnetic field structure
is essential \citep{mariska10} and observations should be complemented with numerical modelling \citep{khomenko13,felipe14}.

%%%%%%%%%%%%%%%%%%%%%%%%%%%%%%%%%%%%%%%%%%%%%%%%%%%%%%%%%%%%%%%%%% 
\section{Theoretical Models and Physical Processes}\label{theory}
%%%%%%%%%%%%%%%%%%%%%%%%%%%%%%%%%%%%%%%%%%%%%%%%%%%%%%%%%%%%%%%%%% 

The observed wave activity is interpreted as the manifestation of the presence of MHD waves. There are three basic MHD wave-types in a uniform plasma of infinite extent and they have properties that make them suitable for wave energy transport and eventual plasma heating. Fast waves transport energy across magnetic surfaces. They resonantly couple to Alfv\'en and slow waves in non-uniform plasmas, which leads to the transfer of energy from large to small spatial scales.  Slow waves are basic in the generation of wave dynamics in the lower solar atmosphere, where the plasma still plays the relevant role in front of the magnetic forces.  They are crucial in non-linear wave transformation processes leading to the development of shocks and their eventual dissipation.  Alfv\'en waves can connect and carry energy between remote regions of the atmosphere. They propagate to large distances along the magnetic field lines. Their energy is not easily dissipated, but their dynamics is characterised by large cross-field gradients in non-uniform plasmas, which leads to the enhancement of viscous and resistive dissipative processes.

None of the observed waves is expected to happen with the properties described in textbooks \citep{goossens03,goedbloed04,priest14}. Pure MHD waves do not exist in the solar atmosphere. The classic description in terms of fast, slow, and Alfv\'en waves treated separately is not accurate enough for the description of the observed dynamics. The highly non-uniform and dynamic nature of the plasma in the solar atmosphere leads to complex wave properties. We identify three levels of complexity in the theoretical and numerical modelling of MHD waves in the solar atmosphere: (1) the analysis of linear waves in simple magnetic configurations enable us to study basic properties of wave trapping and propagation and to understand in detail the workings of individual physical mechanisms. They offer local information on plasma and field properties from seismology analyses, but observational signatures with little potential to be compared to real data; (2)  the analysis of non-linear waves in structured and dynamic plasmas enables us to study wave transformation processes, wave-flow interactions, and the modelling of relevant processes such as shocks waves, non-linear Alfv\'en waves, and Alfv\'en wave turbulence. 
They produce useful synthetic data to extract conclusions from the comparison to observations; (3) global numerical models use prescribed or observed boundaries and drivers as an input on large-scale numerical simulations. They make possible to include almost any physical ingredient, to perform global seismology, and to obtain useful observational signatures and predictions. 

The three approaches offer valuable information and are highly complementary. Results from simple models cannot be directly used to extract solid conclusions, but offer detailed understanding about physical processes that might occur under real Sun conditions. Their benefit is that the mechanism(s) of interest can be studied in isolation and under controlled conditions. On the other extreme, sophisticated simulations enable the direct comparison to observed data, but detailed understanding about the physical processes is lost when many of them are allowed to happen simultaneously in complicated simulations. The level of detailed understanding  will decrease as we increase applicability and vice versa. 

Resonant damping of oscillations is a well studied mechanism for energy transport from large scale transverse motions to localised small scale motions and is due to the inhomogeneity of the medium in the cross-field direction \citep[see][for a review]{goossens06}. The mechanism has been mainly studied in flux tube models \citep{ruderman09,goossens11}, but a cylindrically symmetric flux tube tube is not necessary for the process to operate \citep[see e.g.,][]{rial13}. Resonant absorption is robust in front of model complications, such as non-linear evolution \citep{terradas08c} or multi stranded structure \citep{terradas08b, pascoe11}, and works for both standing \citep{goossens02a,ruderman02} and propagating \citep{pascoe12} waves, producing time and spatial damping of the wave amplitude, respectively. This is an ideal physical process, independent of resistivity in the limit of high Reynolds numbers \citep{poedts91a}. Once in the resonant layer, energy can further scale down to smaller spatial scales by the process of phase mixing \citep{heyvaerts83,hood02}. The stronger the cross-field inhomogeneity, the quicker is the damping, but the slower the small scale creation \citep{terradas06a}. When a given length-scale is reached,
viscous and/or resistive dissipation processes turn on, thus enabling the heating of the plasma. 

To date, two arguments support the mechanism of resonant absorption being operative in the observed waves. First, theory predicts damping time and spatial scales that are compatible with those observed. Second, resonant absorption is a frequency selective process \citep{terradas10}, with low-frequency waves being less damped in front of high frequency waves. A comparison between the outward to inward power ratio measurements by \cite{tomczyk07} using CoMP and the theoretical predictions by \cite{verth10} strongly supports the idea that resonant absorption might be producing the observed in situ energy loss.

Another mechanism studied in detail is Alfv\'en wave turbulence \citep{vanballegooijen11,vanballegooijen14,asgari13}. According to the models, photospheric foot-point motions transported along loops get amplified and reflected at the transition region boundary, thus producing a pattern of counter-propagating perturbations. Their complex interactions lead to the creation of small scales and dissipation. Recent numerical experiments show that, if that is the case, the obtained dissipation rates are able to reproduce chromospheric and coronal heating requirements. The heating rate scales with the magnetic field strength and with the loop length. The modelling of the lower atmosphere is also an important factor. Realistic lower atmosphere modelling favours AC heating in front of DC dissipation.

Observations are now starting to offer support in favour of this process. \cite{demoortel14} analyse CoMP Doppler shifts oscillations and their frequency distribution at both sides of a large trans-equatorial loop system and find that an excess of high frequency power is present at the apex. This is interpreted as being due to low and mid frequency wave energy cascading down because of Alfv\'en wave turbulence. A recent comparison between observed non-thermal velocities with predictions from theoretical models in coronal loops by \cite{asgari14} leads to the conclusion that Alfv\'en wave turbulence is indeed a  strong candidate for explaining how the observed loops are heated.

Significant efforts are being done in the design and application of numerical models, such as the Alfv\'en Wave Solar Model (AWSoM, \citealt{sokolov13,vanderholst14}),  that address the coronal heating problem and the solar wind acceleration from a global perspective. These models incorporate Alfv\'en wave turbulence. The result is a remarkable capacity to reproduce the observed EUV emission and produce solar wind predictions.

Physical processes discussed above are mostly restricted to MHD models. However, the MHD approximation is possibly part
of the difficulty to advance the wave heating question, since the actual heating processes must occur at kinetic scales and hybrid or fully kinetic analyses need to be further developed. The ideal MHD model should fail  when applied to the lower solar atmosphere where partial ionisation effects are important \citep{khomenko12} and effects like ambipolar diffusion, collisions and non-magnetisation become non-negligible. For instance, \cite{vranjes10} have pointed out that, because of the presence of non-magnetised ions in the photosphere, Alfv\'en waves cannot be efficiently generated in, nor travel through, this region. More importantly, the wave energy flux through the photosphere becomes orders of magnitude smaller, compared to the ideal case, when the effects of partial ionisation and collisions are consistently taken into account \citep{vranjes08}. On the other hand, \cite{song11} and \cite{tusong13} have studied Alfv\'en wave propagation and heating through plasma-neutral collisions and find that they can generate sufficient heat, with most of the heat deposited as required at lower altitudes. To clarify these aspects is crucial in view of the recent observations of the chromospheric dynamics with IRIS \citep{depontieu14}.

%%%%%%%%%%%%%%%%%%%%%%%%%%%%%%%%%%%%%%%%%%%%%%%%%%%%%%%%%%%%%%%%%% 
\section{Wave Energy Content, Damping and Dissipation}\label{processes}
%%%%%%%%%%%%%%%%%%%%%%%%%%%%%%%%%%%%%%%%%%%%%%%%%%%%%%%%%%%%%%%%%% 

Our estimates for the energy content on the observed waves are based on simple expressions for bulk Alfv\'en waves in homogeneous plasmas. \cite{goossens13} have shown that the energy flux computed with the well known expression for bulk Alfv\'en waves could overestimate the real flux 
by a factor in the range 10 to 50. \cite{vandoorsselaere14} provide approximations to the energy propagated by kink modes in an ensemble of flux tubes finding a correction factor for the energy in kink waves, compared to the bulk Alfv\'en waves, in terms of the density filling factor.

The full energy balance will be given by Poynting theorem, which relates the amount of electromagnetic energy variation; the fraction of that
energy that propagates with the wave; and the remaining energy that is dissipated. The assessment of these quantities requires the measurement of the spatial and temporal variation of both velocity and magnetic field perturbations. This being a challenge, a first step is the detailed analysis of the flow of energy and its spatial distribution in the waveguide. For resonance absorption in a flux tube model this has been done by \cite{arregui11c}, in the prominence context. The results show that ideal damping is produced by the jump of the radial component of the Poynting vector, which produces the energy flow into the dissipative layer from both sides of the resonance. Once in the resonant layer, energy flows along the field and resistivity, wave fields, and the created currents determine the energy dissipation and its spatial distribution. Quantification and comparison to observed field-aligned brightness variations remain to be done. 

In numerical experiments, all magnetic and plasma perturbation are available. The kinetic, magnetic, and total energy and their spatial and temporal evolution can 
readily be analysed. This is not the case with observations, where the energy contained in the magnetic field is hidden and has to be inferred. The numerical simulations by \cite{demoortel12b}, for transverse wave propagation in a multi-stranded waveguide model, have shown that line of sight integration effects can be important when evaluating the energy content of a wave from only the measured kinetic energy. Also, that a relevant amount of energy that is present in the form of magnetic perturbations in the simulation would be out of observational reach. The ``visible'' energy would only be in the 3\% to 10\% range of the total energy.

As first noted by \cite{lee86},  wave damping is not a synonym for wave dissipation. The two processes might operate at different time and spatial scales and e.g., viscosity should not immediately dissipate the enhanced local oscillations. The matching of the two time-scales is important. Consider for instance resonant damping, phase mixing, and resistive dissipation. A calculation by J. Terradas (personal communication) shows that, under coronal conditions (magnetic Reynolds number of the order of 10$^{12}$) the observational consequence, the attenuation of the motion, might give little information on the concealed physical process, the dissipation of the wave energy. For instance, for a typical coronal loop subject to resonant damping because of a transverse inhomogeneity length scale of $l/R=0.1$ (in units of the tube radius $R$), the damping time over per period is $\tau_{\rm d}/P\sim13$.  The time at which dissipation becomes important, once phase mixing has sufficiently decreased the characteristic transverse spatial scales, is  $\tau_{\rm ra}/P\sim170$. The stronger the non-uniformity, the faster is the damping, $\tau_{\rm d}/P\sim3$ for $l/R=0.5$, but the slower the small scale creation due to phase mixing.  We need therefore the wait more time for resistivity to become important, $\tau_{\rm ra}/P\sim500$.  This means that energy dissipation would only be efficient at very large times, and there would be no heating during the observed oscillations.

The numerical experiment on resonant absorption by \cite{ofman98} analysed the plasma response and the creation of small-scale structure because of the non-linear response of the density, but the assumed energy balance was not entirely self-consistent. More recent numerical experiments by \cite{terradas08b,terradas08c} have analysed 
the first two stages (damping and phase mixing),  in complex models including the non-linearity, but  no 
heating was computed. A detailed numerical study of the full process of wave damping, phase mixing, resistive or viscous dissipation, and plasma heating, 
including the forward modelling of the synthetic plasma response has not yet been undertaken. Such an experiment would prove/disprove the viability of heating by resonant absorption and phase mixing, by
producing spatial and temporal heating profiles to be compared to observations.

%%%%%%%%%%%%%%%%%%%%%%%%%%%%%%%%%%%%%%%%%%%%%%%%%%%%%%%%%%%%%%%%%% 
\section{Confronting Theory and Observations}\label{datatheory}
%%%%%%%%%%%%%%%%%%%%%%%%%%%%%%%%%%%%%%%%%%%%%%%%%%%%%%%%%%%%%%%%%% 

Most of the efforts in the area have focused on the design and use of better instrumentation for data acquisition and analysis and the improvement of theoretical and numerical modelling. The two developments being important, further advance will only come from the development of informative tools for the combination of data and theory to assess the physical conditions and processes that operate in the solar atmosphere. The combination of observed data and theoretical results to infer physical properties and mechanisms is not easy task.  One needs to solve two problems simultaneously. In the forward problem, we prescribe theoretical models and parameters (the causes) and analyse the theoretical wave properties (the consequences).  In the  inverse problem, we try to infer the causes (the unknown physical conditions/mechanisms) from the consequences (the observed wave properties).  The discovery of wave dynamics pervading the solar atmosphere has led to considerable efforts in both forward and inverse modelling.

\subsection{Forward Modelling of Observational Signatures}

The creation of synthetic imaging and spectral observational signatures for direct comparison to data is starting to be used as a means to extract conclusions about the goodness of our wave models \citep[see e.g.,][]{shelyag06,taroyan07,demoortel08,taroyan09,owen09}.

% wave type identification
Forward modelling is essential to solve wave type identification issues. \cite{vandoorsselaere08a} raised the concern as to whether the observed transverse motions should be interpreted as fast kink waves or Alfv\'en waves. The issue was clarified by \cite{goossens09}, who pointed that the observed dynamics is the result of Alfv\'en waves modified by the radial structuring of the plasma density. Their nature being highly Alfv\'enic, these authors refer to them as surface Alfv\'en waves.  Kink and torsional Alfv\'en waves in magnetic flux tubes produce distinct observational signatures \citep{goossens14}. Besides the naming, which is of secondary importance, it is crucial to consider an adequate description of the structures in which the waves propagate, in particular when it comes to seismology and energy budget calculations using the observed waves \citep{goossens12a}.  Observed tilted features in spectrograms have been interpreted as signatures of torsional Alfv\'en waves \citep{depontieu12}. The forward modelling by \cite{goossens14} suggests that tilted features may also arise under the kink wave interpretation, depending among other factors on the line of sight integration.

Another applications of forward modelling techniques include the fitting of wave properties, such as wave periods and amplitudes, from the comparison
of alternative data realisations and observed data \citep{mcintosh11}; the assessment of the wave or flow nature of observed quasi-periodic disturbances from Doppler shifted spectra \citep{depontieu10,tian11,tian12,verwichte10b,kiddie12}; or the analysis of the nature of shock distributions along coronal loops as being due to either non-linear Alfv\'en wave or nano flare heating \citep{antolin08}.

The potential of forward modelling should still be further exploited to compare alternative wave models producing distinct observational consequences and alternative heating mechanisms. Different heating models are known to produce distinct heating profiles along waveguides. For instance, viscous heating is determined by the velocity perturbation profile along waveguides, leading to apex heating, while resistive dissipation is  determined by magnetic field perturbations, thus producing foot-point heating \citep{vandoorsselaere07a}.  The recent development of forward modelling techniques to waves by \cite{antolin13} and \cite{antolin14} show great promise.

\subsection{Seismology Inversion from Observed Wave Properties}

In parallel to theoretical and observational advances, the field of solar atmospheric seismology has emerged.  The term  refers to the study of the physical conditions in solar atmospheric magnetic and plasma structures from the analysis of the observed wave properties. It was first suggested by \cite{uchida70} and \cite{roberts84}, in the coronal context, and by \cite{roberts94}  in the prominence context.  The aim is to increase our knowledge about the complicated structure and dynamics of the solar atmosphere and is based on the fact that the properties of the observed oscillations are determined by the plasma and magnetic field properties.   

Solar atmospheric seismology  has experienced a great advancement  \citep[see][for reviews]{roberts00,demoortel05,nakariakov05,banerjee07,arregui12b,demoortel12}.  This was made possible by the increase in the quantity and quality of wave activity observations and the refinement of theoretical MHD wave models.  

Local seismology, based on the use of simplified models for the magnetic and plasma structuring of tube-like waveguides and observations of the period and damping of standing and propagating waves, have enabled us to obtain information on the magnetic field strength \citep{nakariakov01,vandoorsselaere08}, the coronal density scale height \citep{andries05b,andries09b}, the magnetic field expansion \citep{verth08b}, the Alfv\'en speed \citep{arregui07a,goossens08b} or the longitudinal and cross-field magnetic field and density structuring \citep{verwichte06,verth11,arregui13b,arregui14}. This information is crucial to evaluate the time and spatial scales for wave damping and dissipation processes (Section \ref{processes}). The simplicity of the models considered so far imposes a limited applicability to real Sun conditions. As pointed out by \cite{demoortel09}, current inversion results need to be re-examined in front of the outcome from more involved numerical models, in order to determine their accuracy.  As the number of observed events increases, statistical analyses - both frequentists \citep{verwichte13b} and Bayesian \citep{asensioramos13} - offer valuable wide-ranging information.

Seismology diagnostics will play a key role when it comes to quantify the energy carried out and dissipated by MHD waves, but the area needs to mature by considering models more akin to the structure and dynamics of solar plasmas. Aiming at the inference of magnetic field and plasma properties in spatially extended regions of the atmosphere is another important topic. The available maps with measured wave characteristics by e.g., \cite{tomczyk09} offer a way to perform global seismology, but few global inversion techniques have been devised so far \citep[see e.g.,][]{ballai07,kwon13}.

\subsection{Bayesian Analysis}

Because any inversion process has to be done under conditions in which information is incomplete and uncertain, probabilistic inference is the next step. The solution to a probabilistic inversion problem is given in the form of statements about the plausibility of parameters/models \citep{jaynes03}.  Bayesian analysis then enable us to quantify the degree of belief on these statements, by measuring to which amount their are supported by information on observed data. This is done by application of Bayes' rule \citep{bayes63}, a mathematical theorem that teach us how to combine prior knowledge with the likelihood of obtaining a data realisation as a function of the unknown parameters/models to compute the so-called posterior, which accounts for what can be said about a parameter or model, conditional on data.

The concept can be applied to the problems of parameter inference and model comparison. In parameter inference the posterior is computed for different combinations of parameters and then one marginalises the full posterior to obtain information about the parameter of interest. In model comparison, the ratio of posteriors for alternative models is computed to assess which one better explains observed data. Bayesian analysis is producing successful results in the physical sciences \citep[see e.g., ][]{vontoussaint11} and other areas of space science and astrophysics research, such as cosmology or exoplanet detection \citep{loredo92,gregory05,trotta08} . Applications to solar physics are scarce, but the few applications to the analysis of solar oscillations \citep{marsh08}, the comparison of heating profiles along loops \citep{adamakis10} or to inference and model comparison problems in solar atmospheric seismology \citep{arregui11b,arregui13a} show great promise. 

The main advantages of the Bayesian methodology are that inference is made using self-consistently all the available information in prior knowledge, observed data, and model, additionally obtaining a correct propagation of uncertainty. In model comparison, the method enables to assess quantitatively which model among competing alternative explanations better explains observed data. The development and application of Bayesian techniques can shed light to the problems of wave type identification, wave vs. flow interpretation, and the comparison between alternative wave damping and heating scenarios.

%%%%%%%%%%%%%%%%%%%%%%%%%%%%%%%%%%%%%%%%%%%%%%%%%%%%%%%%%%%%%%%%%% 
\section{Conclusions}\label{conclusions}
%%%%%%%%%%%%%%%%%%%%%%%%%%%%%%%%%%%%%%%%%%%%%%%%%%%%%%%%%%%%%%%%%% 
Physical explanations based on waves were at the forefront of the coronal heating discussion since the problem came to existence. At that time, evidence about the presence of waves and oscillations was absent and the role magnetic fields could play was largely unknown. The solar atmosphere is now one of the best observed and studied astrophysical systems and the presence of waves and oscillations is beyond question. Waves and oscillations are found and analysed in structures with very different physical conditions. Multi-wavelength and multi-scale imaging and spectroscopic observations enable us to measure their properties with increasing precision. 

Wave heating theories remain plausible, but most current models are still too simple to be applicable to the real Sun. Future models should consider the highly structured and dynamic nature of the coronal plasma.  The observed waves act as energy carriers and can transport energy to small scales, where the heating processes occur concealed from observational scrutiny. The theoretical analysis of such kinetic processes should be pursued to derive their large scale observational consequences. Similarly, the consequences of partial ionisation and collisions on the wave properties and their energy dissipation need to be quantified.  These observational consequences need to be fully developed, by forward modelling of synthetic data and comparison to observations. As alternative explanations for the same phenomena arise, there is a need to devise model comparison tools to assess the performance of alternative wave energy transport and dissipation mechanisms in explaining data. Similar methods need to be formulated to confront wave based mechanisms with another explanations, such as mass flows or nanoflare heating. 

Wave energy seems to fill the solar atmosphere. Enough energy is available at the photospheric level and part of it is transmitted above. We do not know exactly how and in which amount. Future observations should concentrate on tracking the flow of energy across different regions of the atmosphere. The amount of energy in magnetic perturbations is hidden. Obtaining estimates of the energy in small scales and dealing with line of sight integration effects represent a challenge for data analysis. The solution is to combine the appropriate modelling with the application of forward modelling and inference tools.

Advances in both theory and observations have placed again waves at the forefront of the discussion. The design and application of tools for comparison between theory and observations is essential. Forward modelling is in its early babyhood and seismology is in its infancy. The development of a self-consistent methodology to combine information from theory and data, such as Bayesian analysis, is still awaiting to be developed and applied to this area. This would enable us to establish what can be plausibly said about the physical conditions and processes operating in the observed wave dynamics, by using all the available information.

%%%%%%%%%%%%%%%%%%%%%%%%%%%%%%%%%%%%%%%%%%%%%%%%%%%%%%%%%%%%%%%%%% 
\begin{acknowledgements} 
This work was supported by a Ram\'on y Cajal Fellowship and project AYA2011-22846 from the Spanish Ministry of Economy and Competitiveness (MINECO).
I am grateful to Ineke De Moortel and Philippa Browning for the invitation to speak at the Royal Society New Approaches in Coronal Heating 
Discussion Meeting. Many thoughts here presented are the result of an exchange of ideas with Jaume Terradas, Ineke De Moortel, Marcel Goossens, Toshifumi Shimizu, and Andr\'es Asensio Ramos, 
over the last years. I received substantial support from Agurtzane Arregi.
\end{acknowledgements} 
%%%%%%%%%%%%%%%%%%%%%%%%%%%%%%%%%%%%%%%%%%%%%%%%%%%%%%%%%%%%%%%%%% 
 
%\bibliographystyle{rspublicnat}

%\bibliography{/Users/arregui/science/bib/arregui}

\begin{thebibliography}{195}
\providecommand{\natexlab}[1]{#1}
\expandafter\ifx\csname urlstyle\endcsname\relax
  \providecommand{\doi}[1]{doi:\discretionary{}{}{}#1}\else
  \providecommand{\doi}{doi:\discretionary{}{}{}\begingroup
  \urlstyle{rm}\Url}\fi

\bibitem[{{Abramov-Maximov} \emph{et~al.}(2011){Abramov-Maximov}, {Gelfreikh},
  {Kobanov}, {Shibasaki} \& {Chupin}}]{abramov11}
{Abramov-Maximov}, V.~E., {Gelfreikh}, G.~B., {Kobanov}, N.~I., {Shibasaki}, K.
  \& {Chupin}, S.~A. 2011 {Multilevel Analysis of Oscillation Motions in Active
  Regions of the Sun}.
\newblock \emph{\solphys}, \textbf{270}, 175--189.
\newblock (\doi{10.1007/s11207-011-9716-7})

\bibitem[{{Adamakis} \emph{et~al.}(2010){Adamakis}, {Walsh} \&
  {Morton-Jones}}]{adamakis10}
{Adamakis}, S., {Walsh}, R.~W. \& {Morton-Jones}, A.~J. 2010 {Constraining
  Coronal Heating: Employing Bayesian Analysis Techniques to Improve the
  Determination of Solar Atmospheric Plasma Parameters}.
\newblock \emph{\solphys}, \textbf{262}, 117--134.
\newblock (\doi{10.1007/s11207-009-9498-3})

\bibitem[{{Alfv\'en}(1941)}]{alfven41}
{Alfv\'en}, H. 1941 On the solar corona.
\newblock \emph{Arkiv f{\"o}r Matematik, Astronomi och Fysik (Band 27A)},
  \textbf{25}, 1--23.

\bibitem[{{Alfv\'en}(1942)}]{alfven42}
{Alfv\'en}, H. 1942 Existence of electromagnetic-hydromagnetic waves.
\newblock \emph{Nature}, \textbf{150}, 405.

\bibitem[{{Alfv{\'e}n}(1947)}]{alfven47}
{Alfv{\'e}n}, H. 1947 {Magneto hydrodynamic waves, and the heating of the solar
  corona}.
\newblock \emph{\mnras}, \textbf{107}, 211.

\bibitem[{{Andries} \emph{et~al.}(2005){Andries}, {Arregui} \&
  {Goossens}}]{andries05b}
{Andries}, J., {Arregui}, I. \& {Goossens}, M. 2005 {Determination of the
  Coronal Density Stratification from the Observation of Harmonic Coronal Loop
  Oscillations}.
\newblock \emph{\apjl}, \textbf{624}, L57--L60.

\bibitem[{{Andries} \emph{et~al.}(2009){Andries}, {van Doorsselaere},
  {Roberts}, {Verth}, {Verwichte} \& {Erd{\'e}lyi}}]{andries09b}
{Andries}, J., {van Doorsselaere}, T., {Roberts}, B., {Verth}, G., {Verwichte},
  E. \& {Erd{\'e}lyi}, R. 2009 {Coronal Seismology by Means of Kink Oscillation
  Overtones}.
\newblock \emph{\ssr}, \textbf{149}, 3--29.
\newblock (\doi{10.1007/s11214-009-9561-2})

\bibitem[{{Antolin} \emph{et~al.}(2008){Antolin}, {Shibata}, {Kudoh}, {Shiota}
  \& {Brooks}}]{antolin08}
{Antolin}, P., {Shibata}, K., {Kudoh}, T., {Shiota}, D. \& {Brooks}, D. 2008
  {Predicting Observational Signatures of Coronal Heating by Alfv{\'e}n Waves
  and Nanoflares}.
\newblock \emph{\apj}, \textbf{688}, 669--682.
\newblock (\doi{10.1086/591998})

\bibitem[{{Antolin} \& {Van Doorsselaere}(2013)}]{antolin13}
{Antolin}, P. \& {Van Doorsselaere}, T. 2013 {Line-of-sight geometrical and
  instrumental resolution effects on intensity perturbations by sausage modes}.
\newblock \emph{\aap}, \textbf{555}, A74.
\newblock (\doi{10.1051/0004-6361/201220784})

\bibitem[{{Antolin} \emph{et~al.}(2014){Antolin}, {Yokoyama} \& {Van
  Doorsselaere}}]{antolin14}
{Antolin}, P., {Yokoyama}, T. \& {Van Doorsselaere}, T. 2014 {Fine Strand-like
  Structure in the Solar Corona from Magnetohydrodynamic Transverse
  Oscillations}.
\newblock \emph{\apjl}, \textbf{787}, L22.
\newblock (\doi{10.1088/2041-8205/787/2/L22})

\bibitem[{{Antonucci} \emph{et~al.}(1984){Antonucci}, {Gabriel} \&
  {Patchett}}]{antonucci84}
{Antonucci}, E., {Gabriel}, A.~H. \& {Patchett}, B.~E. 1984 {Oscillations in
  EUV emission lines during a loop brightening}.
\newblock \emph{\solphys}, \textbf{93}, 85.

\bibitem[{{Arregui}(2012)}]{arregui12b}
{Arregui}, I. 2012 {Inversion of Physical Parameters in Solar Atmospheric
  Seismology}.
\newblock In \emph{Multi-scale dynamical processes in space and astrophysical
  plasmas} (eds M.~P. {Leubner} \& Z.~{V{\"o}r{\"o}s}), p. 159.
\newblock (\doi{10.1007/978-3-642-30442-2-18})

\bibitem[{Arregui \emph{et~al.}(2007)Arregui, Andries, Van~Doorsselaere,
  Goossens \& Poedts}]{arregui07a}
Arregui, I., Andries, J., Van~Doorsselaere, T., Goossens, M. \& Poedts, S. 2007
  {MHD coronal seismology using the period and damping of resonantly damped
  quasi-mode kink oscillations}.
\newblock \emph{Astron. Astrophys.}, \textbf{463}, 333--338.

\bibitem[{{Arregui} \& {Asensio Ramos}(2011)}]{arregui11b}
{Arregui}, I. \& {Asensio Ramos}, A. 2011 {Bayesian Magnetohydrodynamic
  Seismology of Coronal Loops}.
\newblock \emph{\apj}, \textbf{740}, 44.

\bibitem[{{Arregui} \& {Asensio Ramos}(2014)}]{arregui14}
{Arregui}, I. \& {Asensio Ramos}, A. 2014 {Determination of the cross-field
  density structuring in coronal waveguides using the damping of transverse
  waves}.
\newblock \emph{\aap}, \textbf{565}, A78.
\newblock (\doi{10.1051/0004-6361/201423536})

\bibitem[{{Arregui} \emph{et~al.}(2013{\natexlab{\emph{a}}}){Arregui}, {Asensio
  Ramos} \& {D{\'{\i}}az}}]{arregui13a}
{Arregui}, I., {Asensio Ramos}, A. \& {D{\'{\i}}az}, A.~J.
  2013{\natexlab{\emph{a}}} {Bayesian Analysis of Multiple Harmonic
  Oscillations in the Solar Corona}.
\newblock \emph{\apjl}, \textbf{765}, L23.

\bibitem[{{Arregui} \emph{et~al.}(2013{\natexlab{\emph{b}}}){Arregui}, {Asensio
  Ramos} \& {Pascoe}}]{arregui13b}
{Arregui}, I., {Asensio Ramos}, A. \& {Pascoe}, D.~J. 2013{\natexlab{\emph{b}}}
  {Determination of Transverse Density Structuring from Propagating
  Magnetohydrodynamic Waves in the Solar Atmosphere}.
\newblock \emph{\apjl}, \textbf{769}, L34.

\bibitem[{{Arregui} \emph{et~al.}(2012){Arregui}, {Oliver} \&
  {Ballester}}]{arregui12a}
{Arregui}, I., {Oliver}, R. \& {Ballester}, J.~L. 2012 {Prominence
  Oscillations}.
\newblock \emph{Living Reviews in Solar Physics}, \textbf{9}, 2.

\bibitem[{{Arregui} \emph{et~al.}(2011){Arregui}, {Soler}, {Ballester} \&
  {Wright}}]{arregui11c}
{Arregui}, I., {Soler}, R., {Ballester}, J.~L. \& {Wright}, A.~N. 2011
  {Magnetohydrodynamic kink waves in two-dimensional non-uniform prominence
  threads}.
\newblock \emph{\aap}, \textbf{533}, A60.
\newblock (\doi{10.1051/0004-6361/201117477})

\bibitem[{{Aschwanden}(1987)}]{aschwanden87}
{Aschwanden}, M.~J. 1987 {Theory of radio pulsations in coronal loops}.
\newblock \emph{\solphys}, \textbf{111}, 113.

\bibitem[{{Aschwanden}(2003)}]{aschwanden03a}
{Aschwanden}, M.~J. 2003 {Review of Coronal Oscillations - An Observer's View}.
\newblock In \emph{Turbulence, waves and instabilities in the solar plasma,
  nato science series: Ii: Mathematics, physics and chemistry, vol. 124} (eds
  R.~von F\'ay-Siebenb$\ddot{\rm u}$rgen, K.~Petrovay, B.~Roberts \&
  M.~Aschwanden), pp. 215--237. Kluwer Academic Publishers.

\bibitem[{{Aschwanden}(2005)}]{aschwanden05b}
{Aschwanden}, M.~J. 2005 \emph{{Physics of the Solar Corona. An Introduction
  with Problems and Solutions (2nd edition)}}.
\newblock Springer-Praxis.

\bibitem[{{Aschwanden}(2006)}]{aschwanden06}
{Aschwanden}, M.~J. 2006 {Coronal magnetohydrodynamic waves and oscillations:
  observations and quests}.
\newblock \emph{Royal Society of London Philosophical Transactions Series A},
  \textbf{364}, 417--432.

\bibitem[{{Aschwanden} \emph{et~al.}(2002){Aschwanden}, {De Pontieu},
  {Schrijver} \& {Title}}]{aschwanden02}
{Aschwanden}, M.~J., {De Pontieu}, B., {Schrijver}, C.~J. \& {Title}, A.~M.
  2002 {Transverse Oscillations in Coronal Loops Observed with TRACE - II.
  Measurements of Geometric and Physical Parameters}.
\newblock \emph{\solphys}, \textbf{206}, 99.

\bibitem[{{Aschwanden} \emph{et~al.}(1999){Aschwanden}, {Fletcher}, {Schrijver}
  \& {Alexander}}]{aschwanden99}
{Aschwanden}, M.~J., {Fletcher}, L., {Schrijver}, C.~J. \& {Alexander}, D. 1999
  {Coronal Loop Oscillations Observed with the Transition Region and Coronal
  Explorer}.
\newblock \emph{\apj}, \textbf{520}, 880.

\bibitem[{{Asensio Ramos} \& {Arregui}(2013)}]{asensioramos13}
{Asensio Ramos}, A. \& {Arregui}, I. 2013 {Coronal loop physical parameters
  from the analysis of multiple observed transverse oscillations}.
\newblock \emph{\aap}, \textbf{554}, A7.

\bibitem[{{Asgari-Targhi} \emph{et~al.}(2013){Asgari-Targhi}, {van
  Ballegooijen}, {Cranmer} \& {DeLuca}}]{asgari13}
{Asgari-Targhi}, M., {van Ballegooijen}, A.~A., {Cranmer}, S.~R. \& {DeLuca},
  E.~E. 2013 {The Spatial and Temporal Dependence of Coronal Heating by
  Alfv{\'e}n Wave Turbulence}.
\newblock \emph{\apj}, \textbf{773}, 111.
\newblock (\doi{10.1088/0004-637X/773/2/111})

\bibitem[{{Asgari-Targhi} \emph{et~al.}(2014){Asgari-Targhi}, {van
  Ballegooijen} \& {Imada}}]{asgari14}
{Asgari-Targhi}, M., {van Ballegooijen}, A.~A. \& {Imada}, S. 2014 {Comparison
  of Extreme Ultraviolet Imaging Spectrometer Observations of Solar Coronal
  Loops with Alfv{\'e}n Wave Turbulence Models}.
\newblock \emph{\apj}, \textbf{786}, 28.
\newblock (\doi{10.1088/0004-637X/786/1/28})

\bibitem[{{Ballai}(2007)}]{ballai07}
{Ballai}, I. 2007 {Global Coronal Seismology}.
\newblock \emph{\solphys}, \textbf{246}, 177--185.
\newblock (\doi{10.1007/s11207-007-0415-3})

\bibitem[{{Ballester}(2006)}]{ballester06}
{Ballester}, J.~L. 2006 {Recent progress in prominence seismology}.
\newblock \emph{Royal Society of London Philosophical Transactions Series A},
  \textbf{364}, 405--415.

\bibitem[{{Banerjee} \emph{et~al.}(2007){Banerjee}, {Erd{\'e}lyi}, {Oliver} \&
  {O'Shea}}]{banerjee07}
{Banerjee}, D., {Erd{\'e}lyi}, R., {Oliver}, R. \& {O'Shea}, E. 2007 {Present
  and Future Observing Trends in Atmospheric Magnetoseismology}.
\newblock \emph{\solphys}, \textbf{246}, 3--29.
\newblock (\doi{10.1007/s11207-007-9029-z})

\bibitem[{{Banerjee} \emph{et~al.}(2011){Banerjee}, {Gupta} \&
  {Teriaca}}]{banerjee11}
{Banerjee}, D., {Gupta}, G.~R. \& {Teriaca}, L. 2011 {Propagating MHD Waves in
  Coronal Holes}.
\newblock \emph{\ssr}, \textbf{158}, 267--288.
\newblock (\doi{10.1007/s11214-010-9698-z})

\bibitem[{{Banerjee} \emph{et~al.}(2009){Banerjee}, {P{\'e}rez-Su{\'a}rez} \&
  {Doyle}}]{banerjee09}
{Banerjee}, D., {P{\'e}rez-Su{\'a}rez}, D. \& {Doyle}, J.~G. 2009 {Signatures
  of Alfv{\'e}n waves in the polar coronal holes as seen by EIS/Hinode}.
\newblock \emph{\aap}, \textbf{501}, L15--L18.
\newblock (\doi{10.1051/0004-6361/200912242})

\bibitem[{{Bayes} \& {Price}(1763)}]{bayes63}
{Bayes}, M. \& {Price}, M. 1763 {An Essay towards Solving a Problem in the
  Doctrine of Chances. By the Late Rev. Mr. Bayes, F. R. S. Communicated by Mr.
  Price, in a Letter to John Canton, A. M. F. R. S.}
\newblock \emph{Royal Society of London Philosophical Transactions Series I},
  \textbf{53}, 370--418.

\bibitem[{{Bemporad} \& {Abbo}(2012)}]{bemporad12}
{Bemporad}, A. \& {Abbo}, L. 2012 {Spectroscopic Signature of Alfv{\'e}n Waves
  Damping in a Polar Coronal Hole up to 0.4 Solar Radii}.
\newblock \emph{\apj}, \textbf{751}, 110.
\newblock (\doi{10.1088/0004-637X/751/2/110})

\bibitem[{{Berghmans} \& {Clette}(1999)}]{berghmans99}
{Berghmans}, D. \& {Clette}, F. 1999 {Active region EUV transient brightenings
  - First Results by EIT of SOHO JOP80}.
\newblock \emph{\solphys}, \textbf{186}, 207.

\bibitem[{{Biermann}(1946)}]{biermann46}
{Biermann}, L. 1946 Zur deutung der chromospharischen turbulenz und des
  exzesses der uv-strahlung der sonne.
\newblock \emph{Naturwissenschaften}, \textbf{33}, 118.

\bibitem[{{Billings}(1959)}]{billings59}
{Billings}, D.~E. 1959 {Velocity Fields in a Coronal Region with a Possible
  Hydromagnetic Interpretation.}
\newblock \emph{\apj}, \textbf{130}, 215.
\newblock (\doi{10.1086/146710})

\bibitem[{{Bonet} \emph{et~al.}(2008){Bonet}, {M{\'a}rquez}, {S{\'a}nchez
  Almeida}, {Cabello} \& {Domingo}}]{bonet08}
{Bonet}, J.~A., {M{\'a}rquez}, I., {S{\'a}nchez Almeida}, J., {Cabello}, I. \&
  {Domingo}, V. 2008 {Convectively Driven Vortex Flows in the Sun}.
\newblock \emph{\apjl}, \textbf{687}, L131--L134.
\newblock (\doi{10.1086/593329})

\bibitem[{{Brynildsen} \emph{et~al.}(2003){Brynildsen}, {Maltby}, {Brekke},
  {Redvik} \& {Kjeldseth-Moe}}]{brynildsen03}
{Brynildsen}, N., {Maltby}, P., {Brekke}, P., {Redvik}, T. \& {Kjeldseth-Moe},
  O. 2003 {Search for a chromospheric resonator above sunspots}.
\newblock \emph{Advances in Space Research}, \textbf{32}, 1097--1102.
\newblock (\doi{10.1016/S0273-1177(03)00312-0})

\bibitem[{{Cargill} \& {de Moortel}(2011)}]{cargill11}
{Cargill}, P. \& {de Moortel}, I. 2011 {Solar physics: Waves galore}.
\newblock \emph{\nat}, \textbf{475}, 463--464.
\newblock (\doi{10.1038/475463a})

\bibitem[{{Cirtain} \emph{et~al.}(2007){Cirtain}, {Golub}, {Lundquist}, {van
  Ballegooijen}, {Savcheva}, {Shimojo}, {DeLuca}, {Tsuneta}, {Sakao}
  \emph{et~al.}}]{cirtain07}
{Cirtain}, J.~W., {Golub}, L., {Lundquist}, L., {van Ballegooijen}, A.,
  {Savcheva}, A., {Shimojo}, M., {DeLuca}, E., {Tsuneta}, S., {Sakao}, T.
  \emph{et~al.} 2007 {Evidence for Alfv{\'e}n Waves in Solar X-ray Jets}.
\newblock \emph{Science}, \textbf{318}, 1580--.
\newblock (\doi{10.1126/science.1147050})

\bibitem[{{De Moortel}(2005)}]{demoortel05}
{De Moortel}, I. 2005 {An overview of coronal seismology}.
\newblock \emph{Royal Society of London Philosophical Transactions Series A},
  \textbf{363}, 2743--2760.
\newblock (\doi{10.1098/rsta.2005.1665})

\bibitem[{{De Moortel} \& {Bradshaw}(2008)}]{demoortel08}
{De Moortel}, I. \& {Bradshaw}, S.~J. 2008 {Forward Modelling of Coronal
  Intensity Perturbations}.
\newblock \emph{\solphys}, \textbf{252}, 101--119.
\newblock (\doi{10.1007/s11207-008-9238-0})

\bibitem[{{De Moortel} \emph{et~al.}(2002{\natexlab{\emph{a}}}){De Moortel},
  {Hood}, {Ireland} \& {Walsh}}]{demoortel02d}
{De Moortel}, I., {Hood}, A.~W., {Ireland}, J. \& {Walsh}, R.~W.
  2002{\natexlab{\emph{a}}} {Longitudinal intensity oscillations in coronal
  loops observed with TRACE - II. Discussion of Measured Parameters}.
\newblock \emph{\solphys}, \textbf{209}, 89.

\bibitem[{{De Moortel} \emph{et~al.}(2000){De Moortel}, {Ireland} \&
  {Walsh}}]{demoortel00}
{De Moortel}, I., {Ireland}, J. \& {Walsh}, R.~W. 2000 {Observation of
  oscillations in coronal loops}.
\newblock \emph{\aap}, \textbf{355}, L23.

\bibitem[{{De Moortel} \emph{et~al.}(2002{\natexlab{\emph{b}}}){De Moortel},
  {Ireland}, {Walsh} \& {Hood}}]{demoortel02c}
{De Moortel}, I., {Ireland}, J., {Walsh}, R.~W. \& {Hood}, A.~W.
  2002{\natexlab{\emph{b}}} {Longitudinal intensity oscillations in coronal
  loops observed with TRACE - I. Overview of Measured Parameters}.
\newblock \emph{\solphys}, \textbf{209}, 61.

\bibitem[{{De Moortel} \emph{et~al.}(2014){De Moortel}, {McIntosh},
  {Threlfall}, {Bethge} \& {Liu}}]{demoortel14}
{De Moortel}, I., {McIntosh}, S.~W., {Threlfall}, J., {Bethge}, C. \& {Liu}, J.
  2014 {Potential Evidence for the Onset of Alfv{\'e}nic Turbulence in
  Trans-equatorial Coronal Loops}.
\newblock \emph{\apjl}, \textbf{782}, L34.
\newblock (\doi{10.1088/2041-8205/782/2/L34})

\bibitem[{{De Moortel} \& {Nakariakov}(2012)}]{demoortel12}
{De Moortel}, I. \& {Nakariakov}, V.~M. 2012 {Magnetohydrodynamic waves and
  coronal seismology: an overview of recent results}.
\newblock \emph{Royal Society of London Philosophical Transactions Series A},
  \textbf{370}, 3193--3216.
\newblock (\doi{10.1098/rsta.2011.0640})

\bibitem[{{De Moortel} \& {Pascoe}(2009)}]{demoortel09}
{De Moortel}, I. \& {Pascoe}, D.~J. 2009 {Putting Coronal Seismology Estimates
  of the Magnetic Field Strength to the Test}.
\newblock \emph{\apjl}, \textbf{699}, L72--L75.
\newblock (\doi{10.1088/0004-637X/699/2/L72})

\bibitem[{{De Moortel} \& {Pascoe}(2012)}]{demoortel12b}
{De Moortel}, I. \& {Pascoe}, D.~J. 2012 {The Effects of Line-of-sight
  Integration on Multistrand Coronal Loop Oscillations}.
\newblock \emph{\apj}, \textbf{746}, 31.
\newblock (\doi{10.1088/0004-637X/746/1/31})

\bibitem[{{De Pontieu} \emph{et~al.}(2012){De Pontieu}, {Carlsson}, {Rouppe van
  der Voort}, {Rutten}, {Hansteen} \& {Watanabe}}]{depontieu12}
{De Pontieu}, B., {Carlsson}, M., {Rouppe van der Voort}, L.~H.~M., {Rutten},
  R.~J., {Hansteen}, V.~H. \& {Watanabe}, H. 2012 {Ubiquitous Torsional Motions
  in Type II Spicules}.
\newblock \emph{\apjl}, \textbf{752}, L12.
\newblock (\doi{10.1088/2041-8205/752/1/L12})

\bibitem[{{De Pontieu} \& {McIntosh}(2010)}]{depontieu10}
{De Pontieu}, B. \& {McIntosh}, S.~W. 2010 {Quasi-periodic Propagating Signals
  in the Solar Corona: The Signature of Magnetoacoustic Waves or High-velocity
  Upflows?}
\newblock \emph{\apj}, \textbf{722}, 1013--1029.
\newblock (\doi{10.1088/0004-637X/722/2/1013})

\bibitem[{{De Pontieu} \emph{et~al.}(2011){De Pontieu}, {McIntosh}, {Carlsson},
  {Hansteen}, {Tarbell}, {Boerner}, {Martinez-Sykora}, {Schrijver} \&
  {Title}}]{depontieu11}
{De Pontieu}, B., {McIntosh}, S.~W., {Carlsson}, M., {Hansteen}, V.~H.,
  {Tarbell}, T.~D., {Boerner}, P., {Martinez-Sykora}, J., {Schrijver}, C.~J. \&
  {Title}, A.~M. 2011 {The Origins of Hot Plasma in the Solar Corona}.
\newblock \emph{Science}, \textbf{331}, 55--.
\newblock (\doi{10.1126/science.1197738})

\bibitem[{{De Pontieu} \emph{et~al.}(2007){De Pontieu}, {McIntosh}, {Carlsson},
  {Hansteen}, {Tarbell}, {Schrijver}, {Title}, {Shine}, {Tsuneta}
  \emph{et~al.}}]{depontieu07}
{De Pontieu}, B., {McIntosh}, S.~W., {Carlsson}, M., {Hansteen}, V.~H.,
  {Tarbell}, T.~D., {Schrijver}, C.~J., {Title}, A.~M., {Shine}, R.~A.,
  {Tsuneta}, S. \emph{et~al.} 2007 {Chromospheric Alfv{\'e}nic Waves Strong
  Enough to Power the Solar Wind}.
\newblock \emph{Science}, \textbf{318}, 1574--.
\newblock (\doi{10.1126/science.1151747})

\bibitem[{{De Pontieu} \emph{et~al.}(2014){De Pontieu}, {Rouppe van der Voort},
  {McIntosh}, {Pereira}, {Carlsson}, {Hansteen}, {Skogsrud}, {Lemen}, {Title}
  \emph{et~al.}}]{depontieu14}
{De Pontieu}, B., {Rouppe van der Voort}, L., {McIntosh}, S.~W., {Pereira},
  T.~M.~D., {Carlsson}, M., {Hansteen}, V., {Skogsrud}, H., {Lemen}, J.,
  {Title}, A. \emph{et~al.} 2014 {On the prevalence of small-scale twist in the
  solar chromosphere and transition region}.
\newblock \emph{Science}, \textbf{346}, 1255732.
\newblock (\doi{10.1126/science.1255732})

\bibitem[{{DeForest} \& {Gurman}(1998)}]{deforest98}
{DeForest}, C.~E. \& {Gurman}, J.~B. 1998 {Observation of Quasi-periodic
  Compressive Waves in Solar Polar Plumes}.
\newblock \emph{\apj}, \textbf{501}, L217.

\bibitem[{{Deubner} \& {Fleck}(1989)}]{deubner89}
{Deubner}, F.-L. \& {Fleck}, B. 1989 {Dynamics of the solar atmosphere. I -
  Spatio-temporal analysis of waves in the quiet solar atmosphere}.
\newblock \emph{\aap}, \textbf{213}, 423--428.

\bibitem[{{Edl{\'e}n}(1943)}]{edlen43}
{Edl{\'e}n}, B. 1943 {Die Deutung der Emissionslinien im Spektrum der
  Sonnenkorona. Mit 6 Abbildungen.}
\newblock \emph{\zap}, \textbf{22}, 30.

\bibitem[{{Felipe} \emph{et~al.}(2014){Felipe}, {Socas-Navarro} \&
  {Khomenko}}]{felipe14}
{Felipe}, T., {Socas-Navarro}, H. \& {Khomenko}, E. 2014 {Synthetic
  Observations of Wave Propagation in a Sunspot Umbra}.
\newblock \emph{\apj}, \textbf{795}, 9.
\newblock (\doi{10.1088/0004-637X/795/1/9})

\bibitem[{{Freij} \emph{et~al.}(2014){Freij}, {Scullion}, {Nelson}, {Mumford},
  {Wedemeyer} \& {Erd{\'e}lyi}}]{freij14}
{Freij}, N., {Scullion}, E.~M., {Nelson}, C.~J., {Mumford}, S., {Wedemeyer}, S.
  \& {Erd{\'e}lyi}, R. 2014 {The Detection of Upwardly Propagating Waves
  Channeling Energy from the Chromosphere to the Low Corona}.
\newblock \emph{\apj}, \textbf{791}, 61.
\newblock (\doi{10.1088/0004-637X/791/1/61})

\bibitem[{{Goedbloed} \& {Halberstadt}(1994)}]{goedbloed94}
{Goedbloed}, J.~P. \& {Halberstadt}, G. 1994 {Magnetohydrodynamic waves in
  coronal flux tubes}.
\newblock \emph{\aap}, \textbf{286}, 275.

\bibitem[{{Goedbloed} \& {Poedts}(2004)}]{goedbloed04}
{Goedbloed}, J.~P.~H. \& {Poedts}, S. 2004 \emph{{Principles of
  Magnetohydrodynamics}}.
\newblock Cambridge University Press.

\bibitem[{{Golub}(1980)}]{golub80}
{Golub}, L. 1980 {X-ray bright points and the solar cycle}.
\newblock \emph{Royal Society of London Philosophical Transactions Series A},
  \textbf{297}, 595--604.
\newblock (\doi{10.1098/rsta.1980.0235})

\bibitem[{{Goossens}(1991)}]{goossens91}
{Goossens}, M. 1991 {Magnetohydrodynamic Waves and Wave Heating in Nonuniform
  Plasmas}.
\newblock In \emph{Advances in solar system magnetohydrodynamics} (eds A.~W.
  {Hood} \& E.~R. {Priest}), p. 137. Cambridge University Press.

\bibitem[{{Goossens}(2003)}]{goossens03}
{Goossens}, M. 2003 \emph{{An introduction to plasma astrophysics and
  magnetohydrodynamics}}, vol. 294 of \emph{Astrophysics and Space Science
  Library}.
\newblock Kluwer Academic Publishers.

\bibitem[{{Goossens}(2008)}]{goossens08b}
{Goossens}, M. 2008 {Seismology of kink oscillations in coronal loops: Two
  decades of resonant damping}.
\newblock In \emph{Iau symposium} (ed. {R.~Erd{\'e}lyi \&
  C.~A.~Mendoza-Brice{\~n}o}), vol. 247 of \emph{IAU Symposium}, pp. 228--242.
\newblock (\doi{10.1017/S1743921308014920})

\bibitem[{Goossens \emph{et~al.}(2006)Goossens, Andries \&
  Arregui}]{goossens06}
Goossens, M., Andries, J. \& Arregui, I. 2006 Damping of magnetohydrodynamic
  waves by resonant absorption in the solar atmosphere.
\newblock \emph{Royal Society of London Philosophical Transactions Series A},
  \textbf{364}, 433--446.

\bibitem[{{Goossens} \emph{et~al.}(2002){Goossens}, {Andries} \&
  {Aschwanden}}]{goossens02a}
{Goossens}, M., {Andries}, J. \& {Aschwanden}, M.~J. 2002 {Coronal loop
  oscillations. An interpretation in terms of resonant absorption of quasi-mode
  kink oscillations}.
\newblock \emph{\aap}, \textbf{394}, L39.

\bibitem[{{Goossens} \emph{et~al.}(2012){Goossens}, {Andries}, {Soler}, {Van
  Doorsselaere}, {Arregui} \& {Terradas}}]{goossens12a}
{Goossens}, M., {Andries}, J., {Soler}, R., {Van Doorsselaere}, T., {Arregui},
  I. \& {Terradas}, J. 2012 {Surface Alfv{\'e}n Waves in Solar Flux Tubes}.
\newblock \emph{\apj}, \textbf{753}, 111.
\newblock (\doi{10.1088/0004-637X/753/2/111})

\bibitem[{{Goossens} \emph{et~al.}(2011){Goossens}, {Erd{\'e}lyi} \&
  {Ruderman}}]{goossens11}
{Goossens}, M., {Erd{\'e}lyi}, R. \& {Ruderman}, M.~S. 2011 {Resonant MHD Waves
  in the Solar Atmosphere}.
\newblock \emph{\ssr}, \textbf{158}, 289--338.
\newblock (\doi{10.1007/s11214-010-9702-7})

\bibitem[{{Goossens} \emph{et~al.}(2014){Goossens}, {Soler}, {Terradas}, {Van
  Doorsselaere} \& {Verth}}]{goossens14}
{Goossens}, M., {Soler}, R., {Terradas}, J., {Van Doorsselaere}, T. \& {Verth},
  G. 2014 {The Transverse and Rotational Motions of Magnetohydrodynamic Kink
  Waves in the Solar Atmosphere}.
\newblock \emph{\apj}, \textbf{788}, 9.
\newblock (\doi{10.1088/0004-637X/788/1/9})

\bibitem[{{Goossens} \emph{et~al.}(2009){Goossens}, {Terradas}, {Andries},
  {Arregui} \& {Ballester}}]{goossens09}
{Goossens}, M., {Terradas}, J., {Andries}, J., {Arregui}, I. \& {Ballester},
  J.~L. 2009 {On the nature of kink MHD waves in magnetic flux tubes}.
\newblock \emph{\aap}, \textbf{503}, 213--223.
\newblock (\doi{10.1051/0004-6361/200912399})

\bibitem[{{Goossens} \emph{et~al.}(2013){Goossens}, {Van Doorsselaere}, {Soler}
  \& {Verth}}]{goossens13}
{Goossens}, M., {Van Doorsselaere}, T., {Soler}, R. \& {Verth}, G. 2013 {Energy
  Content and Propagation in Transverse Solar Atmospheric Waves}.
\newblock \emph{\apj}, \textbf{768}, 191.
\newblock (\doi{10.1088/0004-637X/768/2/191})

\bibitem[{{Gregory}(2005)}]{gregory05}
{Gregory}, P.~C. 2005 \emph{{Bayesian Logical Data Analysis for the Physical
  Sciences: A Comparative Approach with `Mathematica' Support}}.
\newblock Cambridge University Press.

\bibitem[{{Grotrian}(1939)}]{grotrian39}
{Grotrian}, W. 1939 Zur frage der deutung der linien im spektrum der
  sonnenkorona.
\newblock \emph{Naturwissenschaften}, \textbf{27}, 214.

\bibitem[{{Hahn} \emph{et~al.}(2012){Hahn}, {Landi} \& {Savin}}]{hahn12}
{Hahn}, M., {Landi}, E. \& {Savin}, D.~W. 2012 {Evidence of Wave Damping at Low
  Heights in a Polar Coronal Hole}.
\newblock \emph{\apj}, \textbf{753}, 36.
\newblock (\doi{10.1088/0004-637X/753/1/36})

\bibitem[{{Hahn} \& {Savin}(2013)}]{hahn13}
{Hahn}, M. \& {Savin}, D.~W. 2013 {Observational Quantification of the Energy
  Dissipated by Alfv{\'e}n Waves in a Polar Coronal Hole: Evidence that Waves
  Drive the Fast Solar Wind}.
\newblock \emph{\apj}, \textbf{776}, 78.
\newblock (\doi{10.1088/0004-637X/776/2/78})

\bibitem[{Hahn \& Savin(2014)}]{hahn14}
Hahn, M. \& Savin, D.~W. 2014 Evidence for wave heating of the quiet-sun
  corona.
\newblock \emph{The Astrophysical Journal}, \textbf{795}(2), 111.

\bibitem[{{Heyvaerts} \& {Priest}(1983)}]{heyvaerts83}
{Heyvaerts}, J. \& {Priest}, E.~R. 1983 {Coronal heating by phase-mixed shear
  Alfven waves}.
\newblock \emph{\aap}, \textbf{117}, 220.

\bibitem[{{Hollweg}(1978)}]{hollweg78}
{Hollweg}, J.~V. 1978 {Alfven waves in the solar atmosphere}.
\newblock \emph{\solphys}, \textbf{56}, 305--333.
\newblock (\doi{10.1007/BF00152474})

\bibitem[{{Hollweg} \& {Sterling}(1984)}]{hollweg84}
{Hollweg}, J.~V. \& {Sterling}, A.~C. 1984 {Resonant heating - an
  interpretation of coronal loop data}.
\newblock \emph{\apjl}, \textbf{282}, L31--L33.
\newblock (\doi{10.1086/184298})

\bibitem[{{Hood}(2010)}]{hood10}
{Hood}, A.~W. 2010 {Coronal Heating}.
\newblock In \emph{Lecture notes in physics, berlin springer verlag} (eds
  P.~J.~V. {Garcia} \& J.~M. {Ferreira}), vol. 793 of \emph{Lecture Notes in
  Physics, Berlin Springer Verlag}, p. 109.
\newblock (\doi{10.1007/978-3-642-02289-04})

\bibitem[{{Hood} \emph{et~al.}(2002){Hood}, {Brooks} \& {Wright}}]{hood02}
{Hood}, A.~W., {Brooks}, S.~J. \& {Wright}, A.~N. 2002 {Coronal heating by the
  phase mixing of individual pulses propagating in coronal holes}.
\newblock \emph{Royal Society of London Proceedings Series A}, \textbf{458},
  2307.
\newblock (\doi{10.1098/rspa.2002.0959})

\bibitem[{{Hood} \emph{et~al.}(1997){Hood}, {Gonzalez-Delgado} \&
  {Ireland}}]{hood97}
{Hood}, A.~W., {Gonzalez-Delgado}, D. \& {Ireland}, J. 1997 {Heating of coronal
  loops by phase-mixing.}
\newblock \emph{\aap}, \textbf{324}, 11--14.

\bibitem[{{Ionson}(1978)}]{ionson78}
{Ionson}, J.~A. 1978 {Resonant absorption of Alfvenic surface waves and the
  heating of solar coronal loops}.
\newblock \emph{\apj}, \textbf{226}, 650--673.

\bibitem[{{Jaynes}(2003)}]{jaynes03}
{Jaynes}, E.~T. 2003 \emph{Probability theory: The logic of science}.
\newblock Cambridge University Press.

\bibitem[{{Jess} \emph{et~al.}(2009){Jess}, {Mathioudakis}, {Erd{\'e}lyi},
  {Crockett}, {Keenan} \& {Christian}}]{jess09}
{Jess}, D.~B., {Mathioudakis}, M., {Erd{\'e}lyi}, R., {Crockett}, P.~J.,
  {Keenan}, F.~P. \& {Christian}, D.~J. 2009 {Alfv{\'e}n Waves in the Lower
  Solar Atmosphere}.
\newblock \emph{Science}, \textbf{323}, 1582--.
\newblock (\doi{10.1126/science.1168680})

\bibitem[{{Khomenko} \& {Calvo Santamaria}(2013)}]{khomenko13}
{Khomenko}, E. \& {Calvo Santamaria}, I. 2013 {Magnetohydrodynamic waves driven
  by p-modes}.
\newblock \emph{Journal of Physics Conference Series}, \textbf{440}(1), 012048.
\newblock (\doi{10.1088/1742-6596/440/1/012048})

\bibitem[{{Khomenko} \& {Collados}(2012)}]{khomenko12}
{Khomenko}, E. \& {Collados}, M. 2012 {Heating of the Magnetized Solar
  Chromosphere by Partial Ionization Effects}.
\newblock \emph{\apj}, \textbf{747}, 87.
\newblock (\doi{10.1088/0004-637X/747/2/87})

\bibitem[{{Kiddie} \emph{et~al.}(2012){Kiddie}, {De Moortel}, {Del Zanna},
  {McIntosh} \& {Whittaker}}]{kiddie12}
{Kiddie}, G., {De Moortel}, I., {Del Zanna}, G., {McIntosh}, S.~W. \&
  {Whittaker}, I. 2012 {Propagating Disturbances in Coronal Loops: A Detailed
  Analysis of Propagation Speeds}.
\newblock \emph{\solphys}, \textbf{279}, 427--452.
\newblock (\doi{10.1007/s11207-012-0042-5})

\bibitem[{{Kliem} \emph{et~al.}(2002){Kliem}, {Dammasch}, {Curdt} \&
  {Wilhelm}}]{kliem02}
{Kliem}, B., {Dammasch}, I.~E., {Curdt}, W. \& {Wilhelm}, K. 2002 {Correlated
  Dynamics of Hot and Cool Plasmas in the Main Phase of a Solar Flare}.
\newblock \emph{\apjl}, \textbf{568}, L61--L65.
\newblock (\doi{10.1086/340136})

\bibitem[{{Klimchuk}(2006)}]{klimchuk06}
{Klimchuk}, J.~A. 2006 {On Solving the Coronal Heating Problem}.
\newblock \emph{\solphys}, \textbf{234}, 41--77.
\newblock (\doi{10.1007/s11207-006-0055-z})

\bibitem[{{Kuperus}(1969)}]{kuperus69}
{Kuperus}, M. 1969 {The Heating of the Solar Corona}.
\newblock \emph{\ssr}, \textbf{9}, 713--739.
\newblock (\doi{10.1007/BF00174033})

\bibitem[{{Kuperus} \emph{et~al.}(1981){Kuperus}, {Ionson} \&
  {Spicer}}]{kuperus81}
{Kuperus}, M., {Ionson}, J.~A. \& {Spicer}, D.~S. 1981 {On the theory of
  coronal heating mechanisms}.
\newblock \emph{\araa}, \textbf{19}, 7--40.
\newblock (\doi{10.1146/annurev.aa.19.090181.000255})

\bibitem[{{Kuridze} \emph{et~al.}(2013){Kuridze}, {Verth}, {Mathioudakis},
  {Erd{\'e}lyi}, {Jess}, {Morton}, {Christian} \& {Keenan}}]{kuridze13}
{Kuridze}, D., {Verth}, G., {Mathioudakis}, M., {Erd{\'e}lyi}, R., {Jess},
  D.~B., {Morton}, R.~J., {Christian}, D.~J. \& {Keenan}, F.~P. 2013
  {Characteristics of Transverse Waves in Chromospheric Mottles}.
\newblock \emph{\apj}, \textbf{779}, 82.
\newblock (\doi{10.1088/0004-637X/779/1/82})

\bibitem[{{Kwon} \emph{et~al.}(2013){Kwon}, {Kramar}, {Wang}, {Ofman},
  {Davila}, {Chae} \& {Zhang}}]{kwon13}
{Kwon}, R.-Y., {Kramar}, M., {Wang}, T., {Ofman}, L., {Davila}, J.~M., {Chae},
  J. \& {Zhang}, J. 2013 {Global Coronal Seismology in the Extended Solar
  Corona through Fast Magnetosonic Waves Observed by STEREO SECCHI COR1}.
\newblock \emph{\apj}, \textbf{776}, 55.
\newblock (\doi{10.1088/0004-637X/776/1/55})

\bibitem[{{Lee} \& {Roberts}(1986)}]{lee86}
{Lee}, M.~A. \& {Roberts}, B. 1986 {On the behavior of hydromagnetic surface
  waves}.
\newblock \emph{\apj}, \textbf{301}, 430--439.

\bibitem[{{Loredo}(1992)}]{loredo92}
{Loredo}, T.~J. 1992 {Promise of Bayesian inference for astrophysics.}
\newblock In \emph{Statistical challenges in modern astronomy} (eds E.~D.
  {Feigelson} \& G.~J. {Babu}), pp. 275--297.

\bibitem[{{Mariska} \& {Muglach}(2010)}]{mariska10}
{Mariska}, J.~T. \& {Muglach}, K. 2010 {Doppler-shift, Intensity, and Density
  Oscillations Observed with the Extreme Ultraviolet Imaging Spectrometer on
  Hinode}.
\newblock \emph{\apj}, \textbf{713}, 573--583.
\newblock (\doi{10.1088/0004-637X/713/1/573})

\bibitem[{{Marsh} \emph{et~al.}(2008){Marsh}, {Ireland} \& {Kucera}}]{marsh08}
{Marsh}, M.~S., {Ireland}, J. \& {Kucera}, T. 2008 {Bayesian Analysis of Solar
  Oscillations}.
\newblock \emph{\apj}, \textbf{681}, 672--679.
\newblock (\doi{10.1086/588751})

\bibitem[{{Marsh} \& {Walsh}(2006)}]{marsh06}
{Marsh}, M.~S. \& {Walsh}, R.~W. 2006 {p-Mode Propagation through the
  Transition Region into the Solar Corona. I. Observations}.
\newblock \emph{\apj}, \textbf{643}, 540--548.
\newblock (\doi{10.1086/501450})

\bibitem[{{Mathioudakis} \emph{et~al.}(2013){Mathioudakis}, {Jess} \&
  {Erd{\'e}lyi}}]{mathioudakis13}
{Mathioudakis}, M., {Jess}, D.~B. \& {Erd{\'e}lyi}, R. 2013 {Alfv{\'e}n Waves
  in the Solar Atmosphere. From Theory to Observations}.
\newblock \emph{\ssr}, \textbf{175}, 1--27.
\newblock (\doi{10.1007/s11214-012-9944-7})

\bibitem[{{McIntosh} \emph{et~al.}(2011){McIntosh}, {de Pontieu}, {Carlsson},
  {Hansteen}, {Boerner} \& {Goossens}}]{mcintosh11}
{McIntosh}, S.~W., {de Pontieu}, B., {Carlsson}, M., {Hansteen}, V., {Boerner},
  P. \& {Goossens}, M. 2011 {Alfv{\'e}nic waves with sufficient energy to power
  the quiet solar corona and fast solar wind}.
\newblock \emph{\nat}, \textbf{475}, 477--480.
\newblock (\doi{10.1038/nature10235})

\bibitem[{{McIntosh} \emph{et~al.}(2012){McIntosh}, {Tian}, {Sechler} \& {De
  Pontieu}}]{mcintosh12}
{McIntosh}, S.~W., {Tian}, H., {Sechler}, M. \& {De Pontieu}, B. 2012 {On the
  Doppler Velocity of Emission Line Profiles Formed in the ''Coronal
  Contraflow'' that Is the Chromosphere-Corona Mass Cycle}.
\newblock \emph{\apj}, \textbf{749}, 60.
\newblock (\doi{10.1088/0004-637X/749/1/60})

\bibitem[{{Morton}(2014)}]{morton14}
{Morton}, R.~J. 2014 {Magneto-seismological insights into the penumbral
  chromosphere and evidence for wave damping in spicules}.
\newblock \emph{\aap}, \textbf{566}, A90.
\newblock (\doi{10.1051/0004-6361/201423718})

\bibitem[{{Morton} \& {McLaughlin}(2013)}]{morton13a}
{Morton}, R.~J. \& {McLaughlin}, J.~A. 2013 {Hi-C and AIA observations of
  transverse magnetohydrodynamic waves in active regions}.
\newblock \emph{\aap}, \textbf{553}, L10.
\newblock (\doi{10.1051/0004-6361/201321465})

\bibitem[{{Morton} \emph{et~al.}(2013){Morton}, {Verth}, {Fedun}, {Shelyag} \&
  {Erd{\'e}lyi}}]{morton13b}
{Morton}, R.~J., {Verth}, G., {Fedun}, V., {Shelyag}, S. \& {Erd{\'e}lyi}, R.
  2013 {Evidence for the Photospheric Excitation of Incompressible
  Chromospheric Waves}.
\newblock \emph{\apj}, \textbf{768}, 17.
\newblock (\doi{10.1088/0004-637X/768/1/17})

\bibitem[{{Morton} \emph{et~al.}(2012){Morton}, {Verth}, {Jess}, {Kuridze},
  {Ruderman}, {Mathioudakis} \& {Erd{\'e}lyi}}]{morton12}
{Morton}, R.~J., {Verth}, G., {Jess}, D.~B., {Kuridze}, D., {Ruderman}, M.~S.,
  {Mathioudakis}, M. \& {Erd{\'e}lyi}, R. 2012 {Observations of ubiquitous
  compressive waves in the Sun's chromosphere}.
\newblock \emph{Nature Communications}, \textbf{3}, 1315.
\newblock (\doi{10.1038/ncomms2324})

\bibitem[{{Nakariakov} \& {Ofman}(2001)}]{nakariakov01}
{Nakariakov}, V.~M. \& {Ofman}, L. 2001 {Determination of the coronal magnetic
  field by coronal loop oscillations}.
\newblock \emph{\aap}, \textbf{372}, L53.

\bibitem[{Nakariakov \emph{et~al.}(1999)Nakariakov, Ofman, DeLuca, Roberts \&
  Davila}]{nakariakov99}
Nakariakov, V.~M., Ofman, L., DeLuca, E.~E., Roberts, B. \& Davila, J.~M. 1999
  Trace observations of damped coronal loop oscillations: implications for
  coronal heating.
\newblock \emph{Science}, \textbf{285}, 862--864.
\newblock (\doi{10.1126/science.285.5429.862})

\bibitem[{{Nakariakov} \& {Verwichte}(2005)}]{nakariakov05}
{Nakariakov}, V.~M. \& {Verwichte}, E. 2005 {Coronal Waves and Oscillations}.
\newblock \emph{Living Reviews in Solar Physics}, \textbf{2}, 3--+.

\bibitem[{{Narain} \& {Ulmschneider}(1996)}]{Narain96}
{Narain}, U. \& {Ulmschneider}, P. 1996 {Chromospheric and Coronal Heating
  Mechanisms II}.
\newblock \emph{\ssr}, \textbf{75}, 453--509.
\newblock (\doi{10.1007/BF00833341})

\bibitem[{{Nightingale} \emph{et~al.}(1999){Nightingale}, {Aschwanden} \&
  {Hurlburt}}]{nightingale99}
{Nightingale}, R.~W., {Aschwanden}, M.~J. \& {Hurlburt}, N.~E. 1999 {Time
  Variability of EUV Brightenings in Coronal Loops Observed with TRACE}.
\newblock \emph{\solphys}, \textbf{190}, 249.

\bibitem[{{Ofman} \emph{et~al.}(1995){Ofman}, {Davila} \&
  {Steinolfson}}]{ofman95}
{Ofman}, L., {Davila}, J.~M. \& {Steinolfson}, R.~S. 1995 {Coronal heating by
  the resonant absorption of Alfven waves: Wavenumber scaling laws.}
\newblock \emph{\apj}, \textbf{444}, 471--477.
\newblock (\doi{10.1086/175621})

\bibitem[{{Ofman} \emph{et~al.}(1998){Ofman}, {Klimchuk} \& {Davila}}]{ofman98}
{Ofman}, L., {Klimchuk}, J.~A. \& {Davila}, J.~M. 1998 {A Self-consistent Model
  for the Resonant Heating of Coronal Loops: The Effects of Coupling with the
  Chromosphere}.
\newblock \emph{\apj}, \textbf{493}, 474--479.
\newblock (\doi{10.1086/305109})

\bibitem[{{Ofman} \emph{et~al.}(1997){Ofman}, {Romoli}, {Poletto}, {Noci} \&
  {Kohl}}]{ofman97}
{Ofman}, L., {Romoli}, M., {Poletto}, G., {Noci}, G. \& {Kohl}, J.~L. 1997
  {Ultraviolet Coronagraph Spectrometer Observations of Density Fluctuations in
  the Solar Wind}.
\newblock \emph{\apj}, \textbf{491}, L111.

\bibitem[{{Ofman} \emph{et~al.}(2000){Ofman}, {Romoli}, {Poletto}, {Noci} \&
  {Kohl}}]{ofman00a}
{Ofman}, L., {Romoli}, M., {Poletto}, G., {Noci}, G. \& {Kohl}, J.~L. 2000
  {UVCS WLC Observations of Compressional Waves in the South Polar Coronal
  Hole}.
\newblock \emph{\apj}, \textbf{529}, 592.

\bibitem[{Okamoto \emph{et~al.}(2007)Okamoto, Tsuneta, Berger, Ichimoto,
  Katsukawa, Lites, Nagata, Shibata, Shimizu \emph{et~al.}}]{okamoto07}
Okamoto, T., Tsuneta, S., Berger, T., Ichimoto, K., Katsukawa, Y., Lites, B.,
  Nagata, S., Shibata, K., Shimizu, T. \emph{et~al.} 2007 Coronal transverse
  magnetohydrodynamic waves in a solar prominence.
\newblock \emph{Science}, \textbf{318}, 1577--1580.
\newblock (\doi{10.1126/science.1145447})

\bibitem[{{Okamoto} \& {De Pontieu}(2011)}]{okamoto11}
{Okamoto}, T.~J. \& {De Pontieu}, B. 2011 {Propagating Waves Along Spicules}.
\newblock \emph{\apjl}, \textbf{736}, L24.
\newblock (\doi{10.1088/2041-8205/736/2/L24})

\bibitem[{{O'Shea} \emph{et~al.}(2002){O'Shea}, {Muglach} \& {Fleck}}]{oshea02}
{O'Shea}, E., {Muglach}, K. \& {Fleck}, B. 2002 {Oscillations above sunspots:
  Evidence for propagating waves?}
\newblock \emph{\aap}, \textbf{387}, 642--664.
\newblock (\doi{10.1051/0004-6361:20020375})

\bibitem[{{Owen} \emph{et~al.}(2009){Owen}, {De Moortel} \& {Hood}}]{owen09}
{Owen}, N.~R., {De Moortel}, I. \& {Hood}, A.~W. 2009 {Forward modelling to
  determine the observational signatures of propagating slow waves for TRACE,
  SoHO/CDS, and Hinode/EIS}.
\newblock \emph{\aap}, \textbf{494}, 339--353.
\newblock (\doi{10.1051/0004-6361:200810828})

\bibitem[{{Parker}(1963)}]{parker63}
{Parker}, E.~N. 1963 {Comments on Coronal Heating.}
\newblock In \emph{The solar corona} (ed. J.~W. {Evans}), vol.~16 of \emph{IAU
  Symposium}, p.~11.

\bibitem[{{Parnell} \& {De Moortel}(2012)}]{parnell12}
{Parnell}, C.~E. \& {De Moortel}, I. 2012 {A contemporary view of coronal
  heating}.
\newblock \emph{Royal Society of London Philosophical Transactions Series A},
  \textbf{370}, 3217--3240.

\bibitem[{{Pascoe} \emph{et~al.}(2012){Pascoe}, {Hood}, {de Moortel} \&
  {Wright}}]{pascoe12}
{Pascoe}, D.~J., {Hood}, A.~W., {de Moortel}, I. \& {Wright}, A.~N. 2012
  {Spatial damping of propagating kink waves due to mode coupling}.
\newblock \emph{\aap}, \textbf{539}, A37.
\newblock (\doi{10.1051/0004-6361/201117979})

\bibitem[{{Pascoe} \emph{et~al.}(2013){Pascoe}, {Hood}, {De Moortel} \&
  {Wright}}]{pascoe13}
{Pascoe}, D.~J., {Hood}, A.~W., {De Moortel}, I. \& {Wright}, A.~N. 2013
  {Damping of kink waves by mode coupling. II. Parametric study and
  seismology}.
\newblock \emph{\aap}, \textbf{551}, A40.
\newblock (\doi{10.1051/0004-6361/201220620})

\bibitem[{{Pascoe} \emph{et~al.}(2011){Pascoe}, {Wright} \& {De
  Moortel}}]{pascoe11}
{Pascoe}, D.~J., {Wright}, A.~N. \& {De Moortel}, I. 2011 {Propagating Coupled
  Alfv{\'e}n and Kink Oscillations in an Arbitrary Inhomogeneous Corona}.
\newblock \emph{\apj}, \textbf{731}, 73.
\newblock (\doi{10.1088/0004-637X/731/1/73})

\bibitem[{{Pereira} \emph{et~al.}(2014){Pereira}, {De Pontieu}, {Carlsson},
  {Hansteen}, {Tarbell}, {Lemen}, {Title}, {Boerner}, {Hurlburt}
  \emph{et~al.}}]{pereira14}
{Pereira}, T.~M.~D., {De Pontieu}, B., {Carlsson}, M., {Hansteen}, V.,
  {Tarbell}, T.~D., {Lemen}, J., {Title}, A., {Boerner}, P., {Hurlburt}, N.
  \emph{et~al.} 2014 {An Interface Region Imaging Spectrograph First View on
  Solar Spicules}.
\newblock \emph{\apjl}, \textbf{792}, L15.
\newblock (\doi{10.1088/2041-8205/792/1/L15})

\bibitem[{{Peter} \& {Dwivedi}(2014)}]{peter14}
{Peter}, H. \& {Dwivedi}, B.~N. 2014 {Discovery of the Sun's million-degree hot
  corona}.
\newblock \emph{Frontiers in Astronomy and Space Sciences, Vol.~1, id.~2},
  \textbf{1}, 2.
\newblock (\doi{10.3389/fspas.2014.00002})

\bibitem[{{Peter} \emph{et~al.}(2004){Peter}, {Gudiksen} \&
  {Nordlund}}]{peter04}
{Peter}, H., {Gudiksen}, B.~V. \& {Nordlund}, {\AA}. 2004 {Coronal Heating
  through Braiding of Magnetic Field Lines}.
\newblock \emph{\apjl}, \textbf{617}, L85--L88.
\newblock (\doi{10.1086/427168})

\bibitem[{{Peter} \& {Judge}(1999)}]{peter99}
{Peter}, H. \& {Judge}, P.~G. 1999 {On the Doppler Shifts of Solar Ultraviolet
  Emission Lines}.
\newblock \emph{\apj}, \textbf{522}, 1148--1166.
\newblock (\doi{10.1086/307672})

\bibitem[{{Poedts} \emph{et~al.}(1989){Poedts}, {Goossens} \&
  {Kerner}}]{poedts89}
{Poedts}, S., {Goossens}, M. \& {Kerner}, W. 1989 {Numerical simulation of
  coronal heating by resonant absorption of Alfven waves}.
\newblock \emph{\solphys}, \textbf{123}, 83--115.
\newblock (\doi{10.1007/BF00150014})

\bibitem[{{Poedts} \emph{et~al.}(1990){Poedts}, {Goossens} \&
  {Kerner}}]{poedts90}
{Poedts}, S., {Goossens}, M. \& {Kerner}, W. 1990 {On the efficiency of coronal
  loop heating by resonant absorption}.
\newblock \emph{\apj}, \textbf{360}, 279--287.
\newblock (\doi{10.1086/169118})

\bibitem[{{Poedts} \& {Kerner}(1991)}]{poedts91a}
{Poedts}, S. \& {Kerner}, W. 1991 {Ideal quasimodes reviewed in resistive
  magnetohydrodynamics}.
\newblock \emph{\prl}, \textbf{66}, 2871.

\bibitem[{{Priest}(2014)}]{priest14}
{Priest}, E. 2014 \emph{{Magnetohydrodynamics of the Sun}}.
\newblock Cambridge University Press.

\bibitem[{{Rae} \& {Roberts}(1981)}]{rae81}
{Rae}, I.~C. \& {Roberts}, B. 1981 {Surface waves and the heating of the
  corona}.
\newblock \emph{Geophysical and Astrophysical Fluid Dynamics}, \textbf{18},
  197--226.
\newblock (\doi{10.1080/03091928108208836})

\bibitem[{{Reznikova} \emph{et~al.}(2012){Reznikova}, {Shibasaki}, {Sych} \&
  {Nakariakov}}]{reznikova12}
{Reznikova}, V.~E., {Shibasaki}, K., {Sych}, R.~A. \& {Nakariakov}, V.~M. 2012
  {Three-minute Oscillations above Sunspot Umbra Observed with the Solar
  Dynamics Observatory/Atmospheric Imaging Assembly and Nobeyama
  Radioheliograph}.
\newblock \emph{\apj}, \textbf{746}, 119.
\newblock (\doi{10.1088/0004-637X/746/2/119})

\bibitem[{{Rial} \emph{et~al.}(2013){Rial}, {Arregui}, {Terradas}, {Oliver} \&
  {Ballester}}]{rial13}
{Rial}, S., {Arregui}, I., {Terradas}, J., {Oliver}, R. \& {Ballester}, J.~L.
  2013 {Wave Leakage and Resonant Absorption in a Loop Embedded in a Coronal
  Arcade}.
\newblock \emph{\apj}, \textbf{763}, 16.
\newblock (\doi{10.1088/0004-637X/763/1/16})

\bibitem[{{Robbrecht} \emph{et~al.}(2001){Robbrecht}, {Verwichte}, {Berghmans},
  {Hochedez}, {Poedts} \& {Nakariakov}}]{robbrecht01}
{Robbrecht}, E., {Verwichte}, E., {Berghmans}, D., {Hochedez}, J.~F., {Poedts},
  S. \& {Nakariakov}, V.~M. 2001 {Slow magnetoacoustic waves in coronal loops:
  EIT and TRACE}.
\newblock \emph{\aap}, \textbf{370}, 591.

\bibitem[{{Roberts}(2000)}]{roberts00}
{Roberts}, B. 2000 {Waves and oscillations in the corona - (Invited Review)}.
\newblock \emph{\solphys}, \textbf{193}, 139.

\bibitem[{{Roberts} \emph{et~al.}(1984){Roberts}, {Edwin} \&
  {Benz}}]{roberts84}
{Roberts}, B., {Edwin}, P.~M. \& {Benz}, A.~O. 1984 {On coronal oscillations}.
\newblock \emph{\apj}, \textbf{279}, 857.

\bibitem[{{Roberts} \& {Joarder}(1994)}]{roberts94}
{Roberts}, B. \& {Joarder}, P.~S. 1994 {Oscillations in quiescent prominences}.
\newblock In \emph{Advances in solar physics} (ed. {G.~Belvedere, M.~Rodono, \&
  G.~M.~Simnett}), vol. 432 of \emph{Lecture Notes in Physics, Berlin Springer
  Verlag}, pp. 173--178.

\bibitem[{{Rosenberg}(1970)}]{rosenberg70}
{Rosenberg}, H. 1970 {Evidence for MHD Pulsations in the Solar Corona}.
\newblock \emph{\aap}, \textbf{9}, 159.

\bibitem[{{Ruderman} \& {Erd{\'e}lyi}(2009)}]{ruderman09}
{Ruderman}, M.~S. \& {Erd{\'e}lyi}, R. 2009 {Transverse Oscillations of Coronal
  Loops}.
\newblock \emph{\ssr}, \textbf{149}, 199--228.
\newblock (\doi{10.1007/s11214-009-9535-4})

\bibitem[{{Ruderman} \& {Roberts}(2002)}]{ruderman02}
{Ruderman}, M.~S. \& {Roberts}, B. 2002 {The Damping of Coronal Loop
  Oscillations}.
\newblock \emph{\apj}, \textbf{577}, 475.

\bibitem[{{Schrijver} \emph{et~al.}(2002){Schrijver}, {Aschwanden} \&
  {Title}}]{schrijver02}
{Schrijver}, C.~J., {Aschwanden}, M.~J. \& {Title}, A.~M. 2002 {Transverse
  oscillations in coronal loops observed with TRACE - I. An Overview of Events,
  Movies, and a Discussion of Common Properties and Required Conditions}.
\newblock \emph{\solphys}, \textbf{206}, 69.

\bibitem[{{Schwarzschild}(1948)}]{schwarzschild48}
{Schwarzschild}, M. 1948 {On Noise Arising from the Solar Granulation.}
\newblock \emph{\apj}, \textbf{107}, 1.
\newblock (\doi{10.1086/144983})

\bibitem[{{Shelyag} \emph{et~al.}(2006){Shelyag}, {Erd{\'e}lyi} \&
  {Thompson}}]{shelyag06}
{Shelyag}, S., {Erd{\'e}lyi}, R. \& {Thompson}, M.~J. 2006 {Forward Modeling of
  Acoustic Wave Propagation in the Quiet Solar Subphotosphere}.
\newblock \emph{\apj}, \textbf{651}, 576--583.
\newblock (\doi{10.1086/507463})

\bibitem[{Skogsrud \emph{et~al.}(2014)Skogsrud, van~der Voort \&
  Pontieu}]{skogsrud14}
Skogsrud, H., van~der Voort, L.~R. \& Pontieu, B.~D. 2014 On the multi-threaded
  nature of solar spicules.
\newblock \emph{The Astrophysical Journal Letters}, \textbf{795}(1), L23.

\bibitem[{{Sokolov} \emph{et~al.}(2013){Sokolov}, {van der Holst}, {Oran},
  {Downs}, {Roussev}, {Jin}, {Manchester}, {Evans} \& {Gombosi}}]{sokolov13}
{Sokolov}, I.~V., {van der Holst}, B., {Oran}, R., {Downs}, C., {Roussev},
  I.~I., {Jin}, M., {Manchester}, IV, W.~B., {Evans}, R.~M. \& {Gombosi}, T.~I.
  2013 {Magnetohydrodynamic Waves and Coronal Heating: Unifying Empirical and
  MHD Turbulence Models}.
\newblock \emph{\apj}, \textbf{764}, 23.
\newblock (\doi{10.1088/0004-637X/764/1/23})

\bibitem[{{Song} \& {Vasyli{\=u}nas}(2011)}]{song11}
{Song}, P. \& {Vasyli{\=u}nas}, V.~M. 2011 {Heating of the solar atmosphere by
  strong damping of Alfv{\'e}n waves}.
\newblock \emph{Journal of Geophysical Research (Space Physics)}, \textbf{116},
  A09104.
\newblock (\doi{10.1029/2011JA016679})

\bibitem[{{Stein} \& {Leibacher}(1974)}]{stein74}
{Stein}, R.~F. \& {Leibacher}, J. 1974 {Waves in the solar atmosphere}.
\newblock \emph{\araa}, \textbf{12}, 407--435.
\newblock (\doi{10.1146/annurev.aa.12.090174.002203})

\bibitem[{{Sturrock} \& {Uchida}(1981)}]{sturrock81}
{Sturrock}, P.~A. \& {Uchida}, Y. 1981 {Coronal heating by stochastic magnetic
  pumping}.
\newblock \emph{\apj}, \textbf{246}, 331--336.
\newblock (\doi{10.1086/158926})

\bibitem[{{Taroyan} \& {Erd{\'e}lyi}(2009)}]{taroyan09}
{Taroyan}, Y. \& {Erd{\'e}lyi}, R. 2009 {Heating Diagnostics with MHD Waves}.
\newblock \emph{\ssr}, \textbf{149}, 229--254.
\newblock (\doi{10.1007/s11214-009-9506-9})

\bibitem[{{Taroyan} \emph{et~al.}(2007){Taroyan}, {Erd{\'e}lyi}, {Wang} \&
  {Bradshaw}}]{taroyan07}
{Taroyan}, Y., {Erd{\'e}lyi}, R., {Wang}, T.~J. \& {Bradshaw}, S.~J. 2007
  {Forward Modeling of Hot Loop Oscillations Observed by SUMER and SXT}.
\newblock \emph{\apjl}, \textbf{659}, L173--L176.
\newblock (\doi{10.1086/517521})

\bibitem[{{Terradas} \emph{et~al.}(2008{\natexlab{\emph{a}}}){Terradas},
  {Andries}, {Goossens}, {Arregui}, {Oliver} \& {Ballester}}]{terradas08c}
{Terradas}, J., {Andries}, J., {Goossens}, M., {Arregui}, I., {Oliver}, R. \&
  {Ballester}, J.~L. 2008{\natexlab{\emph{a}}} {Nonlinear Instability of Kink
  Oscillations due to Shear Motions}.
\newblock \emph{\apjl}, \textbf{687}, L115--L118.
\newblock (\doi{10.1086/593203})

\bibitem[{{Terradas} \emph{et~al.}(2008{\natexlab{\emph{b}}}){Terradas},
  {Arregui}, {Oliver}, {Ballester}, {Andries} \& {Goossens}}]{terradas08b}
{Terradas}, J., {Arregui}, I., {Oliver}, R., {Ballester}, J.~L., {Andries}, J.
  \& {Goossens}, M. 2008{\natexlab{\emph{b}}} {Resonant Absorption in
  Complicated Plasma Configurations: Applications to Multistranded Coronal Loop
  Oscillations}.
\newblock \emph{\apj}, \textbf{679}, 1611--1620.
\newblock (\doi{10.1086/586733})

\bibitem[{{Terradas} \emph{et~al.}(2010){Terradas}, {Goossens} \&
  {Verth}}]{terradas10}
{Terradas}, J., {Goossens}, M. \& {Verth}, G. 2010 {Selective spatial damping
  of propagating kink waves due to resonant absorption}.
\newblock \emph{\aap}, \textbf{524}, A23.
\newblock (\doi{10.1051/0004-6361/201014845})

\bibitem[{{Terradas} \emph{et~al.}(2006){Terradas}, {Oliver} \&
  {Ballester}}]{terradas06a}
{Terradas}, J., {Oliver}, R. \& {Ballester}, J.~L. 2006 {Damped Coronal Loop
  Oscillations: Time-dependent Results}.
\newblock \emph{\apj}, \textbf{642}, 533--540.
\newblock (\doi{10.1086/500730})

\bibitem[{{Thurgood} \emph{et~al.}(2014){Thurgood}, {Morton} \&
  {McLaughlin}}]{thurgood14}
{Thurgood}, J.~O., {Morton}, R.~J. \& {McLaughlin}, J.~A. 2014 {First Direct
  Measurements of Transverse Waves in Solar Polar Plumes Using SDO/AIA}.
\newblock \emph{\apjl}, \textbf{790}, L2.
\newblock (\doi{10.1088/2041-8205/790/1/L2})

\bibitem[{{Tian} \emph{et~al.}(2011){Tian}, {McIntosh} \& {De
  Pontieu}}]{tian11}
{Tian}, H., {McIntosh}, S.~W. \& {De Pontieu}, B. 2011 {The Spectroscopic
  Signature of Quasi-periodic Upflows in Active Region Timeseries}.
\newblock \emph{\apjl}, \textbf{727}, L37.
\newblock (\doi{10.1088/2041-8205/727/2/L37})

\bibitem[{{Tian} \emph{et~al.}(2012){Tian}, {McIntosh}, {Wang}, {Ofman}, {De
  Pontieu}, {Innes} \& {Peter}}]{tian12}
{Tian}, H., {McIntosh}, S.~W., {Wang}, T., {Ofman}, L., {De Pontieu}, B.,
  {Innes}, D.~E. \& {Peter}, H. 2012 {Persistent Doppler Shift Oscillations
  Observed with Hinode/EIS in the Solar Corona: Spectroscopic Signatures of
  Alfv{\'e}nic Waves and Recurring Upflows}.
\newblock \emph{\apj}, \textbf{759}, 144.
\newblock (\doi{10.1088/0004-637X/759/2/144})

\bibitem[{{Tomczyk} \& {McIntosh}(2009)}]{tomczyk09}
{Tomczyk}, S. \& {McIntosh}, S.~W. 2009 {Time-Distance Seismology of the Solar
  Corona with CoMP}.
\newblock \emph{\apj}, \textbf{697}, 1384--1391.
\newblock (\doi{10.1088/0004-637X/697/2/1384})

\bibitem[{{Tomczyk} \emph{et~al.}(2007){Tomczyk}, {McIntosh}, {Keil}, {Judge},
  {Schad}, {Seeley} \& {Edmondson}}]{tomczyk07}
{Tomczyk}, S., {McIntosh}, S.~W., {Keil}, S.~L., {Judge}, P.~G., {Schad}, T.,
  {Seeley}, D.~H. \& {Edmondson}, J. 2007 {Alfv{\'e}n Waves in the Solar
  Corona}.
\newblock \emph{Science}, \textbf{317}, 1192--.
\newblock (\doi{10.1126/science.1143304})

\bibitem[{{Trotta}(2008)}]{trotta08}
{Trotta}, R. 2008 {Bayes in the sky: Bayesian inference and model selection in
  cosmology}.
\newblock \emph{Contemporary Physics}, \textbf{49}, 71--104.
\newblock (\doi{10.1080/00107510802066753})

\bibitem[{{Trottet} \emph{et~al.}(1979){Trottet}, {Pick} \&
  {Heyvaerts}}]{trottet79}
{Trottet}, G., {Pick}, M. \& {Heyvaerts}, J. 1979 {Gyro-synchrotron modulation
  in the moving type IV bursts}.
\newblock \emph{\aap}, \textbf{79}, 164.

\bibitem[{{Tsubaki}(1977)}]{tsubaki77}
{Tsubaki}, T. 1977 {Periodic oscillations found in coronal velocity fields}.
\newblock \emph{\solphys}, \textbf{51}, 121--130.
\newblock (\doi{10.1007/BF00240450})

\bibitem[{{Tsubaki}(1988)}]{tsubaki88a}
{Tsubaki}, T. 1988 {Observations of periodic oscillations or waves in the solar
  corona and prominences}.
\newblock In \emph{Solar and stellar coronal structure and dynamics} (ed.
  {R.~C.~Altrock}), pp. 140--149.

\bibitem[{{Tu} \& {Song}(2013)}]{tusong13}
{Tu}, J. \& {Song}, P. 2013 {A Study of Alfv{\'e}n Wave Propagation and Heating
  the Chromosphere}.
\newblock \emph{\apj}, \textbf{777}, 53.
\newblock (\doi{10.1088/0004-637X/777/1/53})

\bibitem[{{Uchida}(1970)}]{uchida70}
{Uchida}, Y. 1970 {Diagnosis of Coronal Magnetic Structure by Flare-Associated
  Hydromagnetic Disturbances}.
\newblock \emph{\pasj}, \textbf{22}, 341.

\bibitem[{{van Ballegooijen}(1986)}]{vanballegooijen86}
{van Ballegooijen}, A.~A. 1986 {Cascade of magnetic energy as a mechanism of
  coronal heating}.
\newblock \emph{\apj}, \textbf{311}, 1001--1014.
\newblock (\doi{10.1086/164837})

\bibitem[{{van Ballegooijen} \emph{et~al.}(2014){van Ballegooijen},
  {Asgari-Targhi} \& {Berger}}]{vanballegooijen14}
{van Ballegooijen}, A.~A., {Asgari-Targhi}, M. \& {Berger}, M.~A. 2014 {On the
  Relationship Between Photospheric Footpoint Motions and Coronal Heating in
  Solar Active Regions}.
\newblock \emph{\apj}, \textbf{787}, 87.
\newblock (\doi{10.1088/0004-637X/787/1/87})

\bibitem[{{van Ballegooijen} \emph{et~al.}(2011){van Ballegooijen},
  {Asgari-Targhi}, {Cranmer} \& {DeLuca}}]{vanballegooijen11}
{van Ballegooijen}, A.~A., {Asgari-Targhi}, M., {Cranmer}, S.~R. \& {DeLuca},
  E.~E. 2011 {Heating of the Solar Chromosphere and Corona by Alfv{\'e}n Wave
  Turbulence}.
\newblock \emph{\apj}, \textbf{736}, 3.
\newblock (\doi{10.1088/0004-637X/736/1/3})

\bibitem[{{van der Holst} \emph{et~al.}(2014){van der Holst}, {Sokolov},
  {Meng}, {Jin}, {Manchester}, {T{\'o}th} \& {Gombosi}}]{vanderholst14}
{van der Holst}, B., {Sokolov}, I.~V., {Meng}, X., {Jin}, M., {Manchester}, IV,
  W.~B., {T{\'o}th}, G. \& {Gombosi}, T.~I. 2014 {Alfv{\'e}n Wave Solar Model
  (AWSoM): Coronal Heating}.
\newblock \emph{\apj}, \textbf{782}, 81.
\newblock (\doi{10.1088/0004-637X/782/2/81})

\bibitem[{{Van Doorsselaere} \emph{et~al.}(2007){Van Doorsselaere}, {Andries}
  \& {Poedts}}]{vandoorsselaere07a}
{Van Doorsselaere}, T., {Andries}, J. \& {Poedts}, S. 2007 {Observational
  evidence favors a resistive wave heating mechanism for coronal loops over a
  viscous phenomenon}.
\newblock \emph{\aap}, \textbf{471}, 311--314.
\newblock (\doi{10.1051/0004-6361:20066658})

\bibitem[{{Van Doorsselaere} \emph{et~al.}(2014){Van Doorsselaere}, {Gijsen},
  {Andries} \& {Verth}}]{vandoorsselaere14}
{Van Doorsselaere}, T., {Gijsen}, S.~E., {Andries}, J. \& {Verth}, G. 2014
  {Energy Propagation by Transverse Waves in Multiple Flux Tube Systems Using
  Filling Factors}.
\newblock \emph{\apj}, \textbf{795}, 18.
\newblock (\doi{10.1088/0004-637X/795/1/18})

\bibitem[{{Van Doorsselaere} \emph{et~al.}(2008{\natexlab{\emph{a}}}){Van
  Doorsselaere}, {Nakariakov} \& {Verwichte}}]{vandoorsselaere08a}
{Van Doorsselaere}, T., {Nakariakov}, V.~M. \& {Verwichte}, E.
  2008{\natexlab{\emph{a}}} {Detection of Waves in the Solar Corona: Kink or
  Alfv{\'e}n?}
\newblock \emph{\apjl}, \textbf{676}, L73--L75.
\newblock (\doi{10.1086/587029})

\bibitem[{{Van Doorsselaere} \emph{et~al.}(2008{\natexlab{\emph{b}}}){Van
  Doorsselaere}, {Nakariakov}, {Young} \& {Verwichte}}]{vandoorsselaere08}
{Van Doorsselaere}, T., {Nakariakov}, V.~M., {Young}, P.~R. \& {Verwichte}, E.
  2008{\natexlab{\emph{b}}} {Coronal magnetic field measurement using loop
  oscillations observed by Hinode/EIS}.
\newblock \emph{\aap}, \textbf{487}, L17--L20.
\newblock (\doi{10.1051/0004-6361:200810186})

\bibitem[{{Vernazza} \emph{et~al.}(1975){Vernazza}, {Foukal}, {Noyes},
  {Reeves}, {Schmahl}, {Timothy}, {Withbroe} \& {Huber}}]{vernazza75}
{Vernazza}, J.~E., {Foukal}, P.~V., {Noyes}, R.~W., {Reeves}, E.~M., {Schmahl},
  E.~J., {Timothy}, J.~G., {Withbroe}, G.~L. \& {Huber}, M.~C.~E. 1975 {Time
  variations in extreme-ultraviolet emission lines and the problem of coronal
  heating}.
\newblock \emph{\apjl}, \textbf{199}, L123.

\bibitem[{{Verth} \emph{et~al.}(2008){Verth}, {Erd{\'e}lyi} \&
  {Jess}}]{verth08b}
{Verth}, G., {Erd{\'e}lyi}, R. \& {Jess}, D.~B. 2008 {Refined
  Magnetoseismological Technique for the Solar Corona}.
\newblock \emph{\apjl}, \textbf{687}, L45--L48.
\newblock (\doi{10.1086/593184})

\bibitem[{{Verth} \emph{et~al.}(2011){Verth}, {Goossens} \& {He}}]{verth11}
{Verth}, G., {Goossens}, M. \& {He}, J.-S. 2011 {Magnetoseismological
  Determination of Magnetic Field and Plasma Density Height Variation in a
  Solar Spicule}.
\newblock \emph{\apjl}, \textbf{733}, L15.
\newblock (\doi{10.1088/2041-8205/733/1/L15})

\bibitem[{{Verth} \emph{et~al.}(2010){Verth}, {Terradas} \&
  {Goossens}}]{verth10}
{Verth}, G., {Terradas}, J. \& {Goossens}, M. 2010 {Observational Evidence of
  Resonantly Damped Propagating Kink Waves in the Solar Corona}.
\newblock \emph{\apjl}, \textbf{718}, L102--L105.
\newblock (\doi{10.1088/2041-8205/718/2/L102})

\bibitem[{Verwichte \emph{et~al.}(2006)Verwichte, Foullon \&
  Nakariakov}]{verwichte06}
Verwichte, E., Foullon, C. \& Nakariakov, V. 2006 Seismology of curved coronal
  loops with vertically polarised transverse oscillations.
\newblock \emph{Astron. Astrophys.}, \textbf{452}, 615--622.

\bibitem[{{Verwichte} \emph{et~al.}(2010){Verwichte}, {Marsh}, {Foullon}, {Van
  Doorsselaere}, {De Moortel}, {Hood} \& {Nakariakov}}]{verwichte10b}
{Verwichte}, E., {Marsh}, M., {Foullon}, C., {Van Doorsselaere}, T., {De
  Moortel}, I., {Hood}, A.~W. \& {Nakariakov}, V.~M. 2010 {Periodic Spectral
  Line Asymmetries in Solar Coronal Structures from Slow Magnetoacoustic
  Waves}.
\newblock \emph{\apjl}, \textbf{724}, L194--L198.
\newblock (\doi{10.1088/2041-8205/724/2/L194})

\bibitem[{{Verwichte} \emph{et~al.}(2013){Verwichte}, {Van Doorsselaere},
  {White} \& {Antolin}}]{verwichte13b}
{Verwichte}, E., {Van Doorsselaere}, T., {White}, R.~S. \& {Antolin}, P. 2013
  {Statistical seismology of transverse waves in the solar corona}.
\newblock \emph{\aap}, \textbf{552}, A138.
\newblock (\doi{10.1051/0004-6361/201220456})

\bibitem[{{von Toussaint}(2011)}]{vontoussaint11}
{von Toussaint}, U. 2011 {Bayesian inference in physics}.
\newblock \emph{Reviews of Modern Physics}, \textbf{83}, 943--999.

\bibitem[{{Vranjes} \& {Poedts}(2010)}]{vranjes10}
{Vranjes}, J. \& {Poedts}, S. 2010 {Drift waves in the corona: heating and
  acceleration of ions at frequencies far below the gyrofrequency}.
\newblock \emph{\mnras}, \textbf{408}, 1835--1839.
\newblock (\doi{10.1111/j.1365-2966.2010.17249.x})

\bibitem[{{Vranjes} \emph{et~al.}(2008){Vranjes}, {Poedts}, {Pandey} \& {de
  Pontieu}}]{vranjes08}
{Vranjes}, J., {Poedts}, S., {Pandey}, B.~P. \& {de Pontieu}, B. 2008 {Energy
  flux of Alfv{\'e}n waves in weakly ionized plasma}.
\newblock \emph{\aap}, \textbf{478}, 553--558.
\newblock (\doi{10.1051/0004-6361:20078274})

\bibitem[{{Wang} \emph{et~al.}(2002){Wang}, {Solanki}, {Curdt}, {Innes} \&
  {Dammasch}}]{wang02a}
{Wang}, T.~J., {Solanki}, S.~K., {Curdt}, W., {Innes}, D.~E. \& {Dammasch},
  I.~E. 2002 {Doppler Shift Oscillations of Hot Solar Coronal Plasma Seen by
  SUMER: A Signature of Loop Oscillations?}
\newblock \emph{\apjl}, \textbf{574}, L101--L104.
\newblock (\doi{10.1086/342189})

\bibitem[{{Wedemeyer-B{\"o}hm} \& {Rouppe van der Voort}(2009)}]{wedemeyer09}
{Wedemeyer-B{\"o}hm}, S. \& {Rouppe van der Voort}, L. 2009 {Small-scale swirl
  events in the quiet Sun chromosphere}.
\newblock \emph{\aap}, \textbf{507}, L9--L12.
\newblock (\doi{10.1051/0004-6361/200913380})

\bibitem[{{Wentzel}(1974)}]{wentzel74}
{Wentzel}, D.~G. 1974 {Coronal heating by Alfven waves}.
\newblock \emph{\solphys}, \textbf{39}, 129--140.
\newblock (\doi{10.1007/BF00154975})

\bibitem[{{Wentzel}(1978)}]{wentzel78}
{Wentzel}, D.~G. 1978 {Heating of the solar corona - A new outlook}.
\newblock \emph{Reviews of Geophysics and Space Physics}, \textbf{16},
  757--761.
\newblock (\doi{10.1029/RG016i004p00757})

\bibitem[{{Wentzel}(1979)}]{wentzel79}
{Wentzel}, D.~G. 1979 {The dissipation of hydromagnetic surface waves}.
\newblock \emph{\apj}, \textbf{233}, 756--764.
\newblock (\doi{10.1086/157437})

\bibitem[{{Withbroe} \& {Noyes}(1977)}]{withbroe77}
{Withbroe}, G.~L. \& {Noyes}, R.~W. 1977 {Mass and energy flow in the solar
  chromosphere and corona}.
\newblock \emph{\araa}, \textbf{15}, 363--387.
\newblock (\doi{10.1146/annurev.aa.15.090177.002051})

\bibitem[{{Zweibel} \& {Yamada}(2009)}]{zweibel09}
{Zweibel}, E.~G. \& {Yamada}, M. 2009 {Magnetic Reconnection in Astrophysical
  and Laboratory Plasmas}.
\newblock \emph{\araa}, \textbf{47}, 291--332.
\newblock (\doi{10.1146/annurev-astro-082708-101726})

\end{thebibliography}

\end{document}